\PassOptionsToPackage{unicode}{hyperref}
\PassOptionsToPackage{hyphens}{url}
\PassOptionsToPackage{dvipsnames,svgnames,x11names}{xcolor}
\documentclass[
  11pt]{article}

\usepackage{amsmath,amssymb}
\usepackage{iftex}
\ifPDFTeX
  \usepackage[T1]{fontenc}
  \usepackage[utf8]{inputenc}
  \usepackage{textcomp} 
\else 
  \usepackage{unicode-math}
  \defaultfontfeatures{Scale=MatchLowercase}
  \defaultfontfeatures[\rmfamily]{Ligatures=TeX,Scale=1}
\fi
\usepackage{lmodern}
\ifPDFTeX\else  
\fi
\IfFileExists{upquote.sty}{\usepackage{upquote}}{}
\IfFileExists{microtype.sty}{
  \usepackage[]{microtype}
  \UseMicrotypeSet[protrusion]{basicmath} 
}{}
\makeatletter
\@ifundefined{KOMAClassName}{
  \IfFileExists{parskip.sty}{%
    \usepackage{parskip}
  }{
    \setlength{\parindent}{0pt}
    \setlength{\parskip}{6pt plus 2pt minus 1pt}}
}{
  \KOMAoptions{parskip=half}}
\makeatother
\usepackage{xcolor}
\ifx\paragraph\undefined\else
  \let\oldparagraph\paragraph
  \renewcommand{\paragraph}[1]{\oldparagraph{#1}\mbox{}}
\fi
\ifx\subparagraph\undefined\else
  \let\oldsubparagraph\subparagraph
  \renewcommand{\subparagraph}[1]{\oldsubparagraph{#1}\mbox{}}
\fi

\usepackage{color}
\usepackage{fancyvrb}

\DefineVerbatimEnvironment{Highlighting}{Verbatim}{commandchars=\\\{\}}
\usepackage{framed}
\definecolor{shadecolor}{RGB}{241,243,245}
\newenvironment{Shaded}{\begin{snugshade}}{\end{snugshade}}

\newcommand{\AttributeTok}[1]{\textcolor[rgb]{0.40,0.45,0.13}{#1}}

\newcommand{\ConstantTok}[1]{\textcolor[rgb]{0.56,0.35,0.01}{#1}}
\newcommand{\ControlFlowTok}[1]{\textcolor[rgb]{0.00,0.23,0.31}{#1}}

\newcommand{\DecValTok}[1]{\textcolor[rgb]{0.68,0.00,0.00}{#1}}
\newcommand{\DocumentationTok}[1]{\textcolor[rgb]{0.37,0.37,0.37}{\textit{#1}}}

\newcommand{\FloatTok}[1]{\textcolor[rgb]{0.68,0.00,0.00}{#1}}
\newcommand{\FunctionTok}[1]{\textcolor[rgb]{0.28,0.35,0.67}{#1}}

\newcommand{\NormalTok}[1]{\textcolor[rgb]{0.00,0.23,0.31}{#1}}

\newcommand{\OtherTok}[1]{\textcolor[rgb]{0.00,0.23,0.31}{#1}}

\newcommand{\SpecialCharTok}[1]{\textcolor[rgb]{0.37,0.37,0.37}{#1}}

\newcommand{\StringTok}[1]{\textcolor[rgb]{0.13,0.47,0.30}{#1}}

\providecommand{\tightlist}{%
  \setlength{\itemsep}{0pt}\setlength{\parskip}{0pt}}\usepackage{longtable,booktabs,array}
\usepackage{calc} 
\usepackage{etoolbox}
\makeatletter
\patchcmd\longtable{\par}{\if@noskipsec\mbox{}\fi\par}{}{}
\makeatother
\IfFileExists{footnotehyper.sty}{\usepackage{footnotehyper}}{\usepackage{footnote}}
\makesavenoteenv{longtable}
\usepackage{graphicx}
\makeatletter
\def\maxwidth{\ifdim\Gin@nat@width>\linewidth\linewidth\else\Gin@nat@width\fi}
\def\maxheight{\ifdim\Gin@nat@height>\textheight\textheight\else\Gin@nat@height\fi}
\makeatother
\setkeys{Gin}{width=\maxwidth,height=\maxheight,keepaspectratio}
\makeatletter
\def\fps@figure{htbp}
\makeatother
\newlength{\cslhangindent}
\newlength{\csllabelwidth}
\newlength{\cslentryspacingunit} 
\newenvironment{CSLReferences}[2] 
 {
  \setlength{\parindent}{0pt}
  \ifodd #1
  \let\oldpar\par
  \def\par{\hangindent=\cslhangindent\oldpar}
  \fi
  \setlength{\parskip}{#2\cslentryspacingunit}
 }%
 {}
\usepackage{calc}

\newcommand{\CSLLeftMargin}[1]{\parbox[t]{\csllabelwidth}{#1}}
\newcommand{\CSLRightInline}[1]{\parbox[t]{\linewidth - \csllabelwidth}{#1}\break}

\usepackage{mathtools}
\usepackage{mathtools}
\makeatletter
\@ifpackageloaded{caption}{}{\usepackage{caption}}
\AtBeginDocument{%
\ifdefined\contentsname
  \renewcommand*\contentsname{Table of contents}
\else
  \newcommand\contentsname{Table of contents}
\fi
\ifdefined\listfigurename
  \renewcommand*\listfigurename{List of Figures}
\else
  \newcommand\listfigurename{List of Figures}
\fi
\ifdefined\listtablename
  \renewcommand*\listtablename{List of Tables}
\else
  \newcommand\listtablename{List of Tables}
\fi
\ifdefined\figurename
  \renewcommand*\figurename{Figure}
\else
  \newcommand\figurename{Figure}
\fi
\ifdefined\tablename
  \renewcommand*\tablename{Table}
\else
  \newcommand\tablename{Table}
\fi
}
\@ifpackageloaded{float}{}{\usepackage{float}}
\floatstyle{ruled}
\@ifundefined{c@chapter}{\newfloat{codelisting}{h}{lop}}{\newfloat{codelisting}{h}{lop}[chapter]}
\floatname{codelisting}{Listing}

\makeatother
\makeatletter
\@ifpackageloaded{caption}{}{\usepackage{caption}}
\@ifpackageloaded{subcaption}{}{\usepackage{subcaption}}
\makeatother
\ifLuaTeX
\fi
\IfFileExists{bookmark.sty}{\usepackage{bookmark}}{\usepackage{hyperref}}
\IfFileExists{xurl.sty}{\usepackage{xurl}}{} 
\hypersetup{
  pdftitle={The `Why' behind including `Y' in your imputation model},
  pdfauthor={Lucy D'Agostino McGowan; Sarah C. Lotspeich; Staci A. Hepler},
  pdfkeywords={Bayesian statistics, missing
data, imputation, regression, statistical applications},
  colorlinks=true,
  linkcolor={blue},
  filecolor={Maroon},
  citecolor={Blue},
  urlcolor={Blue},
  pdfcreator={LaTeX via pandoc}}

\begin{document}

\def\spacingset#1{\renewcommand{\baselinestretch}%
{#1}\small\normalsize} \spacingset{1}


\title{\bf The `Why' behind including `Y' in your imputation model}
\author{
Lucy D'Agostino McGowan\\
Department of Statistical Sciences, Wake Forest University\\
and\\Sarah C. Lotspeich\\
Department of Statistical Sciences, Wake Forest University\\
and\\Staci A. Hepler\\
Department of Statistical Sciences, Wake Forest University\\
}
\maketitle

\bigskip
\bigskip
\begin{abstract}
Missing data is a common challenge when analyzing epidemiological data,
and imputation is often used to address this issue. Here, we investigate
the scenario where a covariate used in an analysis has missingness and
will be imputed. There are recommendations to include the outcome from
the analysis model in the imputation model for missing covariates, but
it is not necessarily clear if this recommendation always holds and why
this is sometimes true. We examine deterministic imputation (i.e.,
single imputation with fixed values) and stochastic imputation (i.e.,
single or multiple imputation with random values) methods and their
implications for estimating the relationship between the imputed
covariate and the outcome. We mathematically demonstrate that including
the outcome variable in imputation models is not just a recommendation
but a requirement to achieve unbiased results when using stochastic
imputation methods. Moreover, we dispel common misconceptions about
deterministic imputation models and demonstrate why the outcome should
not be included in these models. This paper aims to bridge the gap
between imputation in theory and in practice, providing mathematical
derivations to explain common statistical recommendations. We offer a
better understanding of the considerations involved in imputing missing
covariates and emphasize when it is necessary to include the outcome
variable in the imputation model.
\end{abstract}

\noindent%
\vfill

\newpage
\spacingset{1.9} 
\hypertarget{introduction}{%
\section{Introduction}\label{introduction}}

In practice, it is common for researchers and practitioners to encounter
missing data in a variable that will be used as a covariate in a future
analysis. Imputation is often employed to handle this issue. Two common
imputation techniques are (i) \emph{deterministic} imputation, where a
single imputation model is fit and the predicted values that fill in the
missing data are treated as fixed in subsequent analyses, and (ii)
\emph{stochastic} imputation, where the uncertainty in the imputation
modeling process is taken into account (e.g., multiple imputation).
Numerous simulation studies have empirically demonstrated how imputation
performs when used to impute missing covariates (1--6); however, there
have been few that bridge the gap between theory and practice by
explaining the mechanics of imputation mathematically.

Of the accessible mathematical derivations that exist, most focus on
summary statistics of the variable being imputed (e.g., how well the
variance for imputed age represents the true variance of age in the
study had it been fully observed) rather than statistics calculated
after conditioning on the imputed variable (e.g., how well the
relationship between cholesterol and imputed age represents the true
relationship between cholesterol and age in the study). When fitting
epidemiological models, we often find ourselves in the latter position
-- fitting models when the covariates have missing data.

Consequently, there is a need for clear mathematical derivations to
explain the impact of imputation on conditional statistics, like
regression model parameters, rather than summary statistics. For
example, while many have discussed including the outcome in imputation
models for missing covariates (1,4,6--8), it may not be well understood
that including the outcome is not just a recommendation but rather a
\emph{requirement} when using stochastic imputation methods in order to
achieve unbiased results. However, including the outcome when performing
deterministic imputation is neither recommended nor required; in fact,
doing so introduces bias in estimated regression coefficients.

There are misconceptions regarding deterministic imputation methods to
address as well. For instance, variability estimates of the imputed
covariate \emph{itself} are underestimated when single, deterministic
imputation is utilized (e.g., the standard deviation of imputed age is
smaller than the true standard deviation of age in the study) (9--12),
potentially leading to the heuristic that deterministic imputation leads
to underestimates of variation. However, when the imputed covariate is
used in a subsequent analysis model, the variability of its regression
coefficient may be under- \emph{or} over-estimated when the imputed
values are treated as fixed. Additionally, under certain conditions, it
is possible to use a resampling technique such as the bootstrap to
recover the correct variance after single, deterministic imputation.
Thus, single, deterministic imputation may be a reasonable approach to
impute data for a regression model, offering valid parameters and
standard error estimates.

This paper aims to bridge the gap between theory and practice of
imputation. We mathematically derive the quantities required to
calculate conditional statistics found in a subsequent regression model
based on imputed covariates. We focus on the common scenario where a
covariate included in the final outcome model has missingness. This
covariate can be the variable of primary interest (e.g., a treatment) or
one that will be used for adjustment (e.g., age or a comorbidity).

The paper is organized as follows. Section 2.1 begins by introducing an
example where values for a covariate of interest are missing at random.
Section 2.2 introduces mathematical properties of deterministic
imputation. Section 2.3 follows, showing the same properties for
stochastic imputation. Section 3 includes an applied example,
demonstrating how practitioners can implement the recommendations in
practice.

\hypertarget{sec-methods}{%
\section{Methods}\label{sec-methods}}

\hypertarget{setup-and-notation}{%
\subsection{Setup and Notation}\label{setup-and-notation}}

\begin{figure}

{\centering \includegraphics{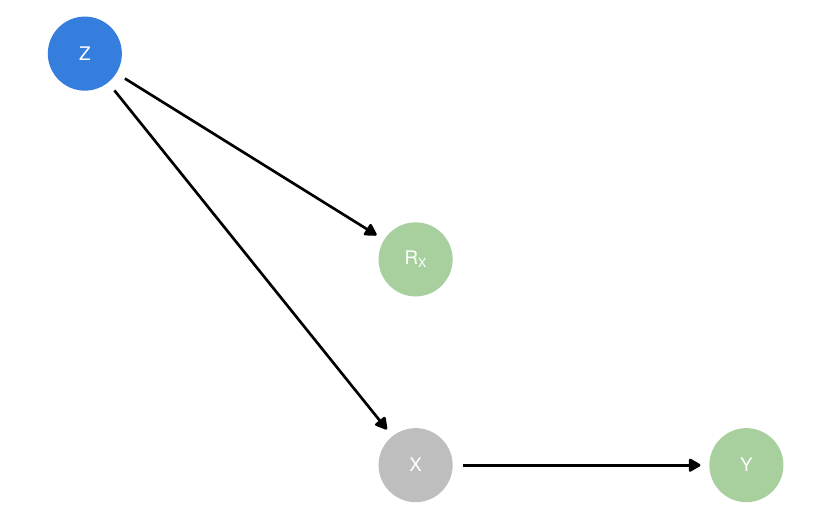}

}

\caption{\label{fig-dag}Missingness directed acyclic graph describing
the relationship between the model outcome, \(Y\), and covariates \(X\)
and \(Z\), and the indicator for missingness in \(X\), \(R_X\).}

\end{figure}

For illustration purposes, we have constructed a simple example. Suppose
we are interested in the relationship between a covariate \(X\) and a
continuous outcome \(Y\). Unfortunately, the observed realizations of
\(X\), \(x\), have some missing values. Throughout, capital letters
denote random variables and lower case letters denote observed
realizations of those variables. We have an additional binary variable,
\(Z\), that is related both to the value of \(X\) and to the chance that
\(X\) is missing. Figure~\ref{fig-dag} shows a \emph{missingness}
directed acyclic graph, or \emph{m-DAG}, as described in (13).

The relationship between \(X\) and \(Z\) can be captured as follows:

\begin{equation}\protect\hypertarget{eq-xz}{}{
X = Z + \varepsilon_X \textrm{, where } Z\sim Bern(0.5) \textrm{ and }\varepsilon_X\sim N(0,1).
}\label{eq-xz}\end{equation}

Thus, when \(Z=1\), \(X\) follows a normal distribution with a mean of 1
and a variance of 1, and when \(Z= 0\), \(X\) follows a normal
distribution with a mean of \(0\) and a variance of 1. Overall, because
\(\Pr(Z=1) = 0.5\), the marginal mean of \(X\) is 0.5 and its variance
is \(1.25\).

Let \(x\) denote the realization or \emph{observed} value of the random
variable \(X\). Let \(R_{X}\) be a missingness indicator for \(x\)
(i.e., \(R_{X}=1\) if \(x\) is missing and \(R_{X}=0\) otherwise) and
\(\mathbf{X}_{obs}\) be a vector representing the subset of
\(\mathbf{X}\) corresponding to the \emph{non-missing} observations,
while \(\mathbf{X}_{mis}\) represents a vector of the missing values.
Thus, \(\mathbf{X}=\mathbf{X}_{obs}\cup \mathbf{X}_{mis}\) We define the
probability mass function of \(R_X\) as

\begin{equation}\protect\hypertarget{eq-miss}{}{
\Pr(R_X = 1|X,Z) = \Pr(R_X = 1|Z) = 0.25(1 - Z) + 0.50(Z).
}\label{eq-miss}\end{equation}

That is, our covariate of interest is missing at random such that if
\(Z=1\) there is a 50\% chance of missingness and if \(Z=0\) the chance
of missingness is 25\%. Since \(\Pr(Z=1) = 0.5\), the unconditional
probability of missing values is \(\Pr(R_X=1)=0.375\).

The analysis model describes the relationship between the covariate,
\(X\), and the outcome, \(Y\). The following equation describes the true
relationship between \(X\) and \(Y\):

\begin{equation}\protect\hypertarget{eq-y}{}{
Y = 2 X + \varepsilon_Y\textrm{, where }\varepsilon_Y \sim N(0,1).
}\label{eq-y}\end{equation}

In other words, the relationship between \(X\) and \(Y\) is linear
through the origin with a slope of 2. Recall that the variance of \(X\)
is 1.25; therefore we know that the true covariance between \(X\) and
\(Y\) is 2.5. Additionally, \(Y\) and \(Z\) are conditionally
independent given \(X\).

The ``complete case'' analysis, that is, the estimation of the
relationship between \(X\) and \(Y\) using only the subset with
non-missing values, \(\mathbf{X}_{obs}\), would be unbiased, since \(Z\)
is not a confounder. Still, the complete case analysis can be less
efficient than an analysis that uses imputation, should the models be
correctly specified. Now, let's explore two methods to overcome the
missingness in the observed \(x\) and estimate the relationship between
\(X\) and \(Y\): deterministic imputation (Section 2.2) and stochastic
imputation (Section 2.3).

\hypertarget{deterministic-imputation}{%
\subsection{Deterministic Imputation}\label{deterministic-imputation}}

Here, by deterministic imputation we mean that we fit a single
imputation model with the non-missing \(x_{obs}\) as the outcome and the
corresponding \(z\) values as the covariate, using regression techniques
(e.g., linear regression). Then, we create predicted values to replace
missing observations based on the fitted imputation model:

\begin{equation}\protect\hypertarget{eq-1}{}{
\widehat{X}_{det} = \widehat\alpha_0 + \widehat\alpha_1z. 
}\label{eq-1}\end{equation}

These predictions are incorporated with non-missing values to construct
the imputed covariate:

\begin{equation}\protect\hypertarget{eq-imp}{}{
X_{imp,det} =(1-R_X)X + R_X\widehat{X}_{det}.
}\label{eq-imp}\end{equation}

To fit the outcome model, we treat the imputed values in a deterministic
fashion (i.e., as fixed). That is, we estimate the relationship between
\(X\) and \(Y\) using \(x_{imp,det}\) in place of \(x\). This procedure
is implemented in the \texttt{mice} R package with the \texttt{method}
argument of the \texttt{mice} function set to \texttt{norm.predict}
(12).

\hypertarget{sec-imp-sum}{%
\subsubsection{Deterministic imputation for univariate summary
statistics}\label{sec-imp-sum}}

Deterministic imputation would be problematic if we were trying to
estimate summary statistics about \(X\) using \(x_{imp,det}\). In this
case, since the covariate is missing at random, the mean of
\(X_{imp,det}\) will be correct, i.e.,
\(\textrm{E}(X_{imp,det}) = \textrm{E}(X)\), as long as our imputation
model in Equation~\ref{eq-1} is correct. However, without additional
correction the variance of \(X_{imp,det}\) will be too small, i.e.,
\(\textrm{Var}(X_{imp,det}) < \textrm{Var}(X)\).

Using the law of total variance, the variance of the deterministically
imputed covariate is

\begin{equation}\protect\hypertarget{eq-def-var}{}{
\begin{aligned}
\textrm{Var}(X_{imp, det}) =& \textrm{E}\{\textrm{Var}(X_{imp, det}|R_X)\} + \textrm{Var}\{\textrm{E}(X_{imp, det}|R_X)\},
\end{aligned}
}\label{eq-def-var}\end{equation}

and its terms can be interpreted as follows. The first term in
Equation~\ref{eq-def-var} is the weighted average of the variance among
the non-missing values and the variance among the imputed values,

\[
\begin{aligned}
\textrm{E}\{\textrm{Var}(X_{imp, det}|R_X)\} =& \textrm{Var}(X|R_X=0)\Pr(R_X=0) + \\&\textrm{Var}(\widehat{X}_{det}|R_X = 1)\Pr(R_X=1), 
\end{aligned}
\]

where the weights depend on the probability of missingness. The second
term in Equation~\ref{eq-def-var} is the squared difference between the
average observed \(X\) and the average imputed \(\widehat{X}_{det}\)
times the variance of the missingness mechanism:

\[
\begin{aligned}
\textrm{Var}&\{\textrm{E}(X_{imp,det}|R_X)\}=\\ &\{\textrm{E}(X|R_X=0)-\textrm{E}(\widehat{X}_{det}|R_X=1)\}^2\Pr(R_X=1)\Pr(R_X=0). \nonumber 
\end{aligned}
\]

Note that if the data were missing completely at random (i.e., the
missingness mechanism did not depend on \(Z\)),
\(\textrm{Var}\{\textrm{E}(X_{imp,det}|R_X)\}\) would equal 0. All
together, the variance of a deterministically imputed covariate can be
summarized in the following annotated formula:

\begin{equation}\protect\hypertarget{eq-var}{}{
\begin{aligned}
\textrm{Var}&(X_{imp,det}) =\\
& \underbrace{\textrm{Var}(X|R_X=0)\Pr(R_X = 0) + \textrm{Var}(\widehat{X}_{det}|R_X = 1)\Pr(R_X = 1)}_{\textrm{E}(\textrm{Var}(X_{imp,det}|R_X))\textrm{ (weighted average of the conditional variances)}} + \\
&\underbrace{\{\textrm{E}(X|R_X=0)-\textrm{E}(\widehat{X}_{det}|R_X=1)\}^2\Pr(R_X=1)\Pr(R_X=0).}_{\textrm{Var}(\textrm{E}(X_{imp,det}|R_X)) \textrm{ (squared difference between average observed and imputed values} \times \textrm{Var}(R_X))}
\end{aligned}
}\label{eq-var}\end{equation}

In fact, the variance of a covariate imputed using \emph{any} method
would have the form in Equation~\ref{eq-var}. In our example,
\(\textrm{Var}(X|R_X=0)\) is going to be slightly smaller than
\(\textrm{Var}(X)\), since more values are missing when \(Z=1\). See
Appendix A for an explanation and derivation. Note that this conditional
variance is the same as the complete case variance of the \(X\) after
removing the missing values. Specifically, \(\textrm{Var}(X|R_X=0)\)
will fall between 1 (the variance among those with the same \(Z\) value)
and 1.25 (the variance for all values of \(X\)). In this scenario, the
complete case variance falls on the upper end of this range, with
\(\textrm{Var}(X|R_X=0)= 1.24\) since \(\Pr(Z=1|R_X=0)=0.6\). Among
those with missing data, the predicted values \(\widehat{X}_{det}\) will
have an expected variance of
\(\textrm{Var}(\widehat{X}_{det}|R_X=1)=0.\overline{22}\), since 66\% of
\(\mathbf{X}_{mis}\) have a value of \(Z=1\). Finally, since 37.5\% of
\(X\) are missing, \(\Pr(R_X = 1) = 0.375\) and the overall expected
variance of \(X_{imp,det}\) is 0.875, as calculated below:

\begin{equation}\protect\hypertarget{eq-var-num}{}{
\begin{aligned}
\textrm{Var}(X_{imp,det}) =&1.24(0.625) + 0.\overline{22}(0.375)+(0.4 - 0.\overline{66})^2(0.625)(0.375)\\
=& 0.875.
\end{aligned}
}\label{eq-var-num}\end{equation}

Note that this quantity is equal to 0.7 times the variance of \(X\)
(\(1.25 \times 0.7 = 0.875\)). Further explanations of these quantities
can be found in Appendix A. Due to this underestimation, in order to
accurately estimate the variance of \(X\) using a deterministically
imputed variable, such as \(X_{imp, det}\), corrections are needed (14).
However, for the scenario in this paper, we are \emph{not} interested in
summary statistics about \(X\) but rather the relationship between \(X\)
and \(Y\).

\hypertarget{sec-imp-sub}{%
\subsubsection{Deterministic imputation for use in a subsequent
model}\label{sec-imp-sub}}

Here, our analysis of interest quantifies the relationship between \(X\)
and \(Y\), for which deterministic imputation yields unbiased results
(15). We fit the outcome model for \(Y\) using \(x_{imp,det}\) in place
of \(x\) as follows:

\begin{equation}\protect\hypertarget{eq-outcome}{}{
\widehat{Y} = \widehat\beta_0 + \widehat\beta_1x_{imp,det}.
}\label{eq-outcome}\end{equation}

Our estimator of interest is \(\widehat\beta_1\), which is unbiased for
\(\beta_1\), the true linear relationship between \(X\) and \(Y\). We
will refer to this as \(\widehat\beta_{1, Y\sim X_{imp,det}}\), with the
notation in the subscript stating the model from which the estimate
comes. This notation will be pertinent as we compare the results from
different imputation methods and modeling choices. In this simple
(unadjusted) linear regression model, we know that
\(\widehat\beta_{1, Y\sim X_{imp,det}} =\widehat{\textrm{Cov}}( X_{imp,det}, Y)/\widehat{\textrm{Var}}(X_{imp,det})\).
From Section~\ref{sec-imp-sum}, \(\textrm{Var}(X_{imp,det})\)
underestimates \(\textrm{Var}(X)\) by a factor of 0.7, and it turns out
that \(\textrm{Cov}(X_{imp,det}, Y)\) \emph{also} underestimates
\(\textrm{Cov}(X, Y)\) by the same multiplicative factor (in this
example, 0.7; see Appendix A for a derivation of a general form of this
factor). The covariance between \(X_{imp,det}\) and \(Y\) is

\begin{equation}\protect\hypertarget{eq-def-cov}{}{
\begin{aligned}
\textrm{Cov}(X_{imp,det}, Y) &= \textrm{E}\{\textrm{Cov}(X_{imp,det}, Y|R_X)\} + \textrm{Cov}\{\textrm{E}(Y|R_X), \textrm{E}(X_{imp,det}|R_X)\}, 
\end{aligned} 
}\label{eq-def-cov}\end{equation}

which is again a function of a weighted average, proportional to the
amount of missingness. The first term in Equation~\ref{eq-def-cov} is
the weighted average of the conditional covariances:

\[
\begin{aligned}
\textrm{E}\{\textrm{Cov}&(X_{imp,det}, Y|R_X)\} =\\& \textrm{Cov}(X, Y|R_X=0)\Pr(R_X=0) + \\&\textrm{Cov}(\widehat{X}_{det}, Y|R_X=1)\Pr(R_X=1), \nonumber 
\end{aligned} 
\]

and the second term is the covariance of the conditional means:

\[
\begin{aligned}
\textrm{Cov}&\{\textrm{E}(Y|R_X), \textrm{E}(X_{imp,det}|R_X)\} =\\ &\textrm{E}(Y|R_X=0)\textrm{E}(X|R_X=0)\Pr(R_X=0)  \\
&+\textrm{E}(Y|R_X=1)\textrm{E}(\widehat{X}_{det}|R_X=1)\Pr(R_X=1) -\textrm{E}(Y)\textrm{E}(X_{imp,det}).
\end{aligned} 
\] Thus, the covariance between an imputed covariate and outcome in a
regression model can be summarized in the following annotated formula:

\begin{equation}\protect\hypertarget{eq-cov}{}{
\begin{aligned}
 \textrm{C}&\textrm{ov}(X_{imp,det}, Y) = 
\\&\underbrace{\textrm{Cov}(X, Y|R_X=0)\Pr(R_X=0) + \textrm{Cov}(\widehat{X}_{det}, Y|R_X=1)\Pr(R_X=1)}_{E\{Cov(X_{imp,det}, Y|R_X)\}\textrm{ (weighted average of the conditional covariances)}}+ \nonumber  \\
& \underbrace{\textrm{E}(Y|R_X=0)\textrm{E}(X|R_X=0)\Pr(R_X=0) +\textrm{E}(Y|R_X=1)\textrm{E}(\widehat{X}_{det}|R_X=1)\Pr(R_X=1) -\textrm{E}(Y)\textrm{E}(X_{imp,det})}_{\textrm{Cov}\{\textrm{E}(Y|R_X), \textrm{E}(X_{imp,det}|R_X)\} \textrm{ (covariance of the conditional means)}}
\end{aligned} 
}\label{eq-cov}\end{equation}

Recall that the true covariance between \(X\) and \(Y\) is 2.5. Since we
did not include \(Y\) in our imputation model (Equation~\ref{eq-1}), the
covariance between \(\widehat{X}_{det}\) and \(Y\) when \(X\) is missing
is smaller
(\(\textrm{Cov}(\widehat{X}_{det}, Y | R_X = 1) = 0.\overline{44}\))
than the covariance between \(X\) and \(Y\). Like the variance in
Section~\ref{sec-imp-sum}, the covariance between the non-missing subset
of \(X\) and \(Y\) will be slightly smaller as well
(\(\textrm{Cov}(X,Y | R_X=0) = 2.48\)). Thus the covariance between
\(X_{imp,det}\) and \(Y\) is 1.75, calculated as follows:

\begin{equation}\protect\hypertarget{eq-cov-num}{}{
\begin{aligned}
\textrm{Cov}(X_{imp,det}, Y) &=2.48(0.625) + 0.\overline{44}(0.375) + 0.8(0.4)(0.625) + 1.\overline{33}(0.\overline{66})(0.375) - 0.5 \\&= 1.75.
\end{aligned}
}\label{eq-cov-num}\end{equation}

Notice that this quantity is 0.7 times the true covariance between \(X\)
and \(Y\) \((2.5\times0.7 = 1.75)\). Recall that the variance of
\(X_{imp,det}\) is off by the same factor.

If we are estimating \(\widehat\beta_{1, Y\sim X_{imp,det}}\), which is
the ratio of the sample covariance and variance, the variance is
underestimated at exactly the same rate that the covariance is
underestimated, making our estimate unbiased for the true relationship
between \(X\) and \(Y\). In other words,
\(\textrm{Cov}(X_{imp,det}, Y)/\textrm{Var}(X_{imp,det}) = 1.75 / 0.875 = 2\),
which is the same ratio we see using the full data:
\(\textrm{Cov}(X, Y)/\textrm{Var}(X) = 2.5 / 1.25 = 2\)
(Figure~\ref{fig-imp-1} in Appendix B). See Appendix A for details and a
derivation of the underestimation factor present in the numerator and
denominator of \(\widehat\beta_{1, Y\sim X_{imp,det}}\). Thus,
deterministic imputation provides unbiased estimates for the
relationship between \(X\) and \(Y\), assuming the imputation model is
correctly specified.

\hypertarget{sec-imp-var}{%
\subsubsection{Variability of model estimates after deterministic
imputation}\label{sec-imp-var}}

One may be concerned about the estimated variance for
\(\widehat\beta_{1, Y\sim X_{imp,det}}\) after performing deterministic
imputation. By default, after plugging in the imputed values \emph{and
treating them as fixed}, the variance for
\(\widehat\beta_{1, Y\sim X_{imp,det}}\) is estimated by:

\begin{equation}\protect\hypertarget{eq-var-beta}{}{
\widehat{\textrm{Var}}\left(\widehat\beta_{1, Y\sim X_{imp,det}}|X_{imp, det}\right)= \left\{\frac{\widehat{\textrm{Var}}(Y)}{\widehat{\textrm{Var}}(X_{imp,det})}-\frac{\widehat{\textrm{Cov}}(X_{imp,det},Y)^2}{\widehat{\textrm{Var}}(X_{imp,det})^2}\right\}\left(\frac{1}{n-2}\right),
}\label{eq-var-beta}\end{equation}

where \(n\) is the total number of observations. Standard regression
software will output estimates of this variance after regressing \(y\)
on \(x_{imp,det}\) after fitting the model in Equation~\ref{eq-outcome}.
In this example, this is approximately equal to \(2.9/(n-2)\) in
expectation. See Appendix A for derivation.

We can compare this model-based estimated variance to several
benchmarks.

  (1) \textbf{Full Cohort Analysis}. First, let's compare
  Equation~\ref{eq-var-beta} to the variance of
  \(\widehat\beta_{1, Y\sim X}\) had no \(X\) been missing and the
  outcome model for \(Y\) using \(X\) from all observations been fit
  instead:

\[
\widehat{\textrm{Var}}\left(\widehat\beta_{1, Y\sim X}\right)= \left\{\frac{\widehat{\textrm{Var}}(Y)}{\widehat{\textrm{Var}}(X)}-\frac{\widehat{\textrm{Cov}}(X,Y)^2}{\widehat{\textrm{Var}}(X)^2}\right\}\left(\frac{1}{n-2}\right). \nonumber
\]

In our example, in expectation
\(\widehat{\textrm{Var}}\left(\widehat\beta_{1, Y\sim X}\right)=0.8/(n-2)\)
(derived in Appendix A). We can rearrange Equation~\ref{eq-var-beta} to
resemble
\(\widehat{\textrm{Var}}\left(\widehat\beta_{1, Y\sim X}\right)\). We
know that both the covariance and the variance of \(X_{imp,det}\) will
be incorrect estimates for the covariance and variance of \(X\)
(Section~\ref{sec-imp-sum} and Section~\ref{sec-imp-sub}), \emph{but} we
know they will be off by the same constant factor, i.e.,

\begin{equation}\protect\hypertarget{eq-cov-omega}{}{
\textrm{Var}(X_{imp,det}) = \textrm{Var}(X)\omega\text{ and }
\textrm{Cov}(X_{imp,det}, Y)=\textrm{Cov}(X,Y)\omega,
}\label{eq-cov-omega}\end{equation}

so their ratio will remain intact. Plugging Equation~\ref{eq-cov-omega}
into Equation~\ref{eq-var-beta} results in:

\begin{equation}\protect\hypertarget{eq-var-beta-2}{}{
\widehat{\textrm{Var}}\left(\widehat\beta_{1, Y\sim X_{imp,det}}|X_{imp,det}\right)=\left\{ \frac{\widehat{\textrm{Var}}(Y)}{\widehat{\textrm{Var}}(X)\omega}-\frac{\widehat{\textrm{Cov}}(X,Y)^2}{\widehat{\textrm{Var}}(X)^2}\right\}\left(\frac{1}{n-2}\right).
}\label{eq-var-beta-2}\end{equation}

When \(\omega<1\) this variance will be greater than the true variance
of \(\widehat\beta_{1, Y\sim X}\) without missing data.

  (2) \textbf{Complete Case Analysis.} Now, let's compare
  Equation~\ref{eq-var-beta} to the variance of
  \(\widehat\beta_{1, Y\sim X|R_X=0}\), that is, the variance from the
  complete case model:

\[
\widehat{\textrm{Var}}\left(\widehat\beta_{1, Y\sim X|R_X=0}\right) = \left\{\frac{\widehat{\textrm{Var}}(Y|R_X=0)}{\widehat{\textrm{Var}}(X|R_X=0)}-\frac{\widehat{\textrm{Cov}}(X,Y|R_X=0)^2}{\widehat{\textrm{Var}}(X|R_X=0)^2}\right\}\left(\frac{1}{n_{obs}-2}\right),
\] where \(n_{obs}\) represents the total number of non-missing values
in \(\mathbf{X}_{obs}\). In our example, the variance of the coefficient
from the complete case model in expectation is approximately
\(0.8/\{n(0.625)-2\}\) (derived in Appendix A). In this example, this is
less than the model-based estimate but greater than the variance of the
parameter given \(X\) had been fully observed.

  (3) \textbf{Analysis with Missing Data (True Sampling Distribution)}.
  Finally, let's compare Equation~\ref{eq-var-beta} to the variance of
  the true sampling distribution of
  \(\widehat\beta_{1, Y\sim X_{imp,det}}\) (the true variance propagates
  the uncertainty of the imputation model). While the variance of
  \(\widehat{\beta}_{1, Y\sim X_{imp,det}}\) using imputed
  \(x_{imp,det}\) and treating the imputed values as fixed
  (Equation~\ref{eq-var-beta}) is larger than the variance of
  \(\widehat\beta_{1, Y\sim X}\) assuming \(X\) had been fully observed
  (Equation~\ref{eq-var-beta-2}), it may over- \emph{or} under-estimate
  the variance of the true sampling distribution of
  \(\widehat\beta_{1, Y\sim X_{imp,det}}\). In this example, the
  variance of the true sampling distribution of
  \(\widehat\beta_{1, Y\sim X_{imp,det}}\) is approximately \(2/(n-2)\)
  in expectation.

For example, if \(n = 102\), on average the \emph{model-based variance}
for the parameter in the analysis with missing data is
\(\widehat{\textrm{Var}}\left(\widehat\beta_{1,Y\sim X_{imp,det}}|X_{imp,det}\right) \approx 0.029\),
the variance for the parameter in the \emph{full cohort analysis} is
\(\widehat{\textrm{Var}}\left(\widehat\beta_{1,Y\sim X}\right) = 0.008\),
the variance for the parameter in the \emph{complete case analysis} is
\(\widehat{\textrm{Var}}\left(\widehat\beta_{1,Y\sim X|R_X=0}\right) \approx 0.013\),
and the \emph{true sampling variance} for the parameter in the analysis
with missing data is
\(\widehat{\textrm{Var}}\left(\widehat\beta_{1,Y\sim X_{imp,det}}\right) \approx 0.02\).
Of note in this example, the variance that appropriately takes into
account the variability in the imputation step (0.02) is greater than
the variance in the complete case analysis (0.013). Thus, in this case,
the complete case analysis would actually be a more efficient approach
compared to deterministic imputation.

If the goal of the analysis is inference, it is important to report the
correct variance, as described in benchmark (3) above, (not using the
default Equation~\ref{eq-var-beta}, which treats the imputed values as
fixed). The correct variance can be estimated by using a resampling
method, such as bootstrap.

\hypertarget{sec-imp-y}{%
\subsubsection{Deterministic imputation using the
outcome}\label{sec-imp-y}}

Let's suppose we confused the common advice to include the outcome in
our imputation model (advice that is meant for stochastic imputation
methods only) and assume the imputation model includes both \(Z\) and
\(Y\) as covariates. Thus, Equation~\ref{eq-1} would be replaced with

\[
\widehat{X}_{det|y}=\widehat\gamma_0 + \widehat\gamma_1z + \widehat\gamma_2 y,
\] and Equation~\ref{eq-imp} would be replaced with

\[
X_{imp,det|y} = (1-R_X)X + R_X\widehat{X}_{det|y}.
\]

Imputing values in this way yields the correct covariance, with
\(\textrm{Cov}(X_{imp,det|y}, Y)= \textrm{Cov}(X,Y)= 2.5\), but
\emph{not} the correct variance, with
\(\textrm{Var}(X_{imp,det|y})=1.175\neq \textrm{Var}(X)=1.25\). The
covariance between \(X_{imp,det|y, Y}\) is correct because the
covariance between \(X_{imp,det|y}\) and \(Y\) when \(X\) is missing is
corrected (replacing the \(0.\overline{44}\) in
Equation~\ref{eq-cov-num} with \(2.\overline{44}\)), with the remainder
of Equation~\ref{eq-cov-num} staying the same. Thus, the variance of the
imputed values among those for whom \(X\) is missing increases
(replacing \(0.\overline{22}\) in Equation~\ref{eq-var-num} with
\(1.0\overline{22}\)) with the remainder of Equation~\ref{eq-var-num}
staying the same, which together brings the
\(\textrm{Var}(X_{imp,det|y})\) to 1.175, rather than the true variance
of 1.25. See Appendix A for further details.

Therefore, deterministic imputation using the outcome, i.e., estimating
the relationship between \(X\) and \(Y\) using the relationship between
\(X_{imp,det|y}\) and \(Y\), will lead to biased estimates
(\(\widehat\beta_{1,Y\sim X_{imp,det|y}}=\frac{2.5}{1.175} \approx 2.1\))
(Figure~\ref{fig-imp-2} in Appendix B).

\hypertarget{sec-stoc}{%
\subsection{Stochastic Imputation}\label{sec-stoc}}

Stochastic imputation is motivated by a Bayesian framework where imputed
values are sampled from a distribution of possibilities, and this is
usually repeated multiple times to create multiple imputed data sets
(16). This approach differs from deterministic imputation, where each
missing value is replaced with a single prediction, implicitly treating
the imputed values as fixed. Let's use Bayesian linear regression to
obtain the predictions. Instead of predicting \(\widehat{X}_{det}\) once
for each missing \(X_{mis}\) in a deterministic fashion, we simulate
multiple values of \(\widehat{X}_{stoc}\) from a posterior predictive
distribution of \(X\), conditional on \(\mathbf{X}_{obs}\) and
\(\mathbf{Z}\).

Suppose we fit the same regression imputation model as in
Equation~\ref{eq-1}, but now we take a Bayesian perspective and specify
noninformative prior distributions for the regression coefficients and
error variance of the imputation model. The imputed data set comes from
simulating a draw from the posterior predictive distribution for the
observations with missingness, or the distribution of \({X}_{stoc}\)
conditional on \(\mathbf{X}_{obs}\) and \(\mathbf{Z}\). The simulated
value is obtained by first drawing \(\dot\sigma^2\) from a scaled
inverse-chi squared distribution with \(n_{obs}-2\) degrees of freedom
and scale parameter equal to the mean squared residual from
Equation~\ref{eq-1}. Then the regression coefficients,
\(\dot{\alpha}_0\) and \(\dot{\alpha}_1\), are simulated jointly from a
bivariate normal distribution with mean equal to the least-squares
estimates (\(\widehat\alpha_0, \widehat\alpha_1\) from
Equation~\ref{eq-1}) and the corresponding variance-covariance matrix.
Lastly, the predicted value, \(\widehat{X}_{stoc}\), is generated by
simulating from a normal distribution with mean
\(\dot{\alpha}_0 + \dot{\alpha}_1 z\) and variance \(\dot\sigma^2\),
which is the distribution that follows from the imputation model

\begin{equation}\protect\hypertarget{eq-stoc-imp}{}{
\widehat{X}_{stoc} = \dot{\alpha}_0 + \dot{\alpha}_1 z + \dot{\varepsilon},
}\label{eq-stoc-imp}\end{equation}

where \(\dot{\varepsilon} \sim N(0, \dot{\sigma}^2)\). In practice, this
process is often repeated to create multiple imputed data sets. This
approach is essentially what is recommended from the earliest multiple
imputation texts (16) and exactly what is implemented in commonly-used
software. For example, this approach is implemented in \texttt{mice}
using the \texttt{norm} method (12). Now the stochastically imputed
covariate values are given by the following:

\[
X_{imp, stoc} = (1-R_X)X + R_X\widehat{X}_{stoc}.
\]

\hypertarget{stochastic-imputation-for-summary-statistics}{%
\subsubsection{Stochastic imputation for summary
statistics}\label{stochastic-imputation-for-summary-statistics}}

This additional variability from simulating parameters for the
imputation model does not change the mean of \(X_{imp,stoc}\). As with
the deterministic imputation, the mean of the imputed covariate will
still be correct (\(0.50\) in our example). However, given the model
specified in Equation~\ref{eq-stoc-imp}, stochastic imputation will
\emph{also} recover the correct variance of \(X\) (14). That is, the
variance of \(X_{imp,stoc}\) will match that of the true variance of
\(X\) (in this case equal to 1.25). Recall, however, that we are not
interested in summary statistics about \(X\) itself, but rather the
relationship between \(X\) and \(Y\), so we also need to estimate the
correct covariance between these variables.

\hypertarget{sec-stoc-imp}{%
\subsubsection{Stochastic imputation for use in a subsequent
model}\label{sec-stoc-imp}}

By recovering the correct variance of \(X\), the desired relationship
between \(X\) and \(Y\) has broken, since
\(\textrm{Cov}(X_{imp,stoc}, Y)\) will still underestimate
\(\textrm{Cov}(X, Y)\). In fact, the covariance between \(X_{imp,stoc}\)
and \(Y\) will be the same as the covariance between \(X_{imp,det}\) and
\(Y\) in Equation~\ref{eq-cov}. See Appendix A for the proof.

Thus, by correcting the variance of \(X_{imp,stoc}\) with this
stochastic process, we are biasing our estimates of the relationship
between \(X\) and \(Y\). Back to our example, if we refit
Equation~\ref{eq-outcome}, plugging in \(x_{imp, stoc}\) rather than
\(x_{imp,det}\), the observed coefficient estimate,
\(\widehat\beta_{1, Y\sim X_{imp, stoc}}\), will be equal to 1.4 in
expectation (Figure~\ref{fig-imp-3} in Appendix B), as the ratio is as
follows:

\[
\widehat\beta_{1, Y\sim X_{imp, stoc}} = \frac{\widehat{\textrm{Cov}}(X_{imp, stoc}, Y)}{\widehat{\textrm{Var}}(X_{imp,stoc})}=\frac{1.75}{1.25}= 1.4 \neq \widehat\beta_{1,Y\sim X}.
\]

\hypertarget{stochastic-imputation-using-the-outcome}{%
\subsubsection{Stochastic imputation using the
outcome}\label{stochastic-imputation-using-the-outcome}}

Fortunately, we can still obtain unbiased \(\widehat\beta_1\) estimates
using stochastic imputation by including the outcome in the imputation
model, which is why the inclusion of the outcome is often recommended
(or we would argue \emph{required}). That is, assume the imputation
model

\[
X_{stoc|y} = \gamma_0 + \gamma_1 Z + \gamma_2 Y + \varepsilon,
\]

where \(\varepsilon \sim N(0, \sigma^2)\). Then, as in
Section~\ref{sec-stoc}, for the observations with missingness, simulate
values \(\widehat{X}_{stoc|y}\) from the posterior predictive
distribution conditional on \(\mathbf{X}_{obs}\), \(\mathbf{Y}\), and
\(\mathbf{Z}\). That is, first simulate values of the regression
coefficients and variance parameter from the posterior distribution,
denoted by \(\dot{\gamma}_0\), \(\dot{\gamma}_1\), \(\dot{\gamma}_2\),
and \(\dot{\sigma}^2\). Then, simulate predicted values,
\(\widehat{X}_{stoc|y}\), from

\begin{equation}\protect\hypertarget{eq-imp-stoc}{}{
\widehat{X}_{stoc|y} = \dot\gamma_0 + \dot\gamma_1z + \dot\gamma_2y + \dot\varepsilon,
}\label{eq-imp-stoc}\end{equation}

and define imputed values as

\[
X_{imp, stoc |y} = (1-R_X)X + R_X\widehat{X}_{stoc|y}.
\]

As demonstrated in Appendix B Figure~\ref{fig-imp-4}, the inclusion of
\(Y\) in the imputation model will recover the appropriate covariance
between \(X_{imp,stoc|y}\) and \(Y\) and the stochastic imputation
process will correctly recover the variance such that
\(\textrm{Var}(X_{imp,stoc|y})\) is equal to \(\textrm{Var}(X)\) as well
(1,4,6,8,16,17). Therefore, unbiased estimates for \(\beta_1\) can be
obtained through stochastic imputation by including the outcome in the
imputation model.

The need to include \(Y\) in the stochastic imputation model is related
to the need for the analysis and imputation models to be ``congenial.''
Despite the fact that it is often completed as a two-step procedure,
stochastic imputation models are inherently imitating a Bayesian process
(see Appendix C for further details). The models are said to be
congenial if there exists a unifying Bayesian model which embeds the
imputation model and the ultimate analysis model which uses the imputed
values (18,19). In other words, in order for the imputation model to be
congenial, it must include all variables that will be in the final
analysis model (including the outcome itself). In fact, not only should
it include all of the variables, but they must be in the same form in
the imputation model as they are in the analysis model. That is, if a
transformation of a covariate is included in the analysis model (e.g., a
quadratic term), the same version of that covariate must be included in
the imputation model (20--22).

\hypertarget{quantifying-variability-after-stochastic-imputation}{%
\subsubsection{Quantifying variability after stochastic
imputation}\label{quantifying-variability-after-stochastic-imputation}}

A single stochastic imputation will result in
\(\textrm{Var}(X_{imp,stoc|Y})=\textrm{Var}(X)\). Likewise,
\(\widehat{\textrm{Var}}\left(\widehat\beta_{1,Y\sim X_{imp,stoc|Y}}\right)\),
i.e., the estimated variance from standard regression software after
fitting the model in Equation~\ref{eq-outcome} (replacing
\(X_{imp, det}\) with \(X_{imp, stoc,|Y}\)), will be equal in
expectation to the variance of the coefficient if \(X\) had been fully
observed, \(\widehat\beta_{1,Y\sim X}\). However, treating the imputed
values as fixed in this way will \emph{underestimate} the true variance
of the sampling distribution of
\(\widehat\beta_{1, Y\sim X_{imp,stoc|Y}}\) (where the true sampling
distribution takes into account the uncertainty of the imputation
model). This underestimation can be corrected by either using the Robins
and Wang asymptotic variance estimator (23), or via multiple imputation,
with the variance calculated using Rubin's Rules (16,24).

\hypertarget{applied-example}{%
\section{Applied Example}\label{applied-example}}

We begin by demonstrating how to fit the above models using the
\texttt{mice} package in R (12). We have simulated a data frame,
\texttt{data}, with three variables: \texttt{x}, \texttt{y}, and
\texttt{z}. Code to generate this data can be seen below.

\spacingset{1}

\begin{Shaded}
\begin{Highlighting}[]
\DocumentationTok{\#\# Generate data for applied example}

\FunctionTok{set.seed}\NormalTok{(}\DecValTok{1}\NormalTok{)}
\NormalTok{n }\OtherTok{\textless{}{-}} \DecValTok{1000}
\NormalTok{z }\OtherTok{\textless{}{-}} \FunctionTok{rbinom}\NormalTok{(n, }\DecValTok{1}\NormalTok{, }\AttributeTok{p =} \FloatTok{0.5}\NormalTok{)}
\NormalTok{x\_full }\OtherTok{\textless{}{-}}\NormalTok{ z }\SpecialCharTok{+} \FunctionTok{rnorm}\NormalTok{(n)}
\NormalTok{y }\OtherTok{\textless{}{-}} \DecValTok{2} \SpecialCharTok{*}\NormalTok{ x\_full }\SpecialCharTok{+} \FunctionTok{rnorm}\NormalTok{(n)}
\NormalTok{r\_x }\OtherTok{\textless{}{-}} \FunctionTok{ifelse}\NormalTok{(z, }\FunctionTok{rbinom}\NormalTok{(n, }\DecValTok{1}\NormalTok{, }\FloatTok{0.5}\NormalTok{), }\FunctionTok{rbinom}\NormalTok{(n, }\DecValTok{1}\NormalTok{, }\FloatTok{0.25}\NormalTok{))}
\NormalTok{x }\OtherTok{\textless{}{-}} \FunctionTok{ifelse}\NormalTok{(r\_x, }\ConstantTok{NA}\NormalTok{, x\_full)}

\NormalTok{data }\OtherTok{\textless{}{-}} \FunctionTok{data.frame}\NormalTok{(}
  \AttributeTok{x =}\NormalTok{ x,}
  \AttributeTok{z =}\NormalTok{ z,}
  \AttributeTok{y =}\NormalTok{ y}
\NormalTok{)}
\end{Highlighting}
\end{Shaded}

\spacingset{1.9}

\hypertarget{deterministic-imputation-1}{%
\subsection{Deterministic Imputation}\label{deterministic-imputation-1}}

We begin by estimating \(X_{imp,det}\) as specified in
Equation~\ref{eq-imp}. We first need to create a prediction matrix. By
default, \texttt{mice} assumes we want to include \texttt{y} in our
imputation model; we need to remove this for deterministic imputation.

\spacingset{1}

\begin{Shaded}
\begin{Highlighting}[]
\DocumentationTok{\#\# Load library}
\FunctionTok{library}\NormalTok{(mice)}


\DocumentationTok{\#\# Create a matrix where we remove y from the imputation model}
\NormalTok{predict\_mat }\OtherTok{\textless{}{-}} \FunctionTok{make.predictorMatrix}\NormalTok{(data)}
\NormalTok{predict\_mat\_noy }\OtherTok{\textless{}{-}}\NormalTok{ predict\_mat}
\NormalTok{predict\_mat\_noy[, }\StringTok{"y"}\NormalTok{] }\OtherTok{\textless{}{-}} \DecValTok{0}
\end{Highlighting}
\end{Shaded}

\spacingset{1.9}

Note that while we are only imputing one variable here, \texttt{x}, if
we had additional variables with missingness, \texttt{mice} would
generalize this process to estimate imputed values for \emph{any}
missing variable, using the remaining variables specified using the
\texttt{predictMatrix} argument, which we explain below. We use the
\texttt{mice} function with the following parameters:

\begin{itemize}
\tightlist
\item
  \texttt{m\ =\ 1} allows us to complete a \emph{single} imputation
\item
  \texttt{method\ =\ "norm.predict"} allows us to apply
  \emph{deterministic} imputation
\item
  \texttt{predictMatrix\ =\ predict\_mat\_noy} replaces the default
  prediction matrix with one that does not include \texttt{y} in the
  imputation model(s)
\end{itemize}

Finally, the complete data set can be extracted using the
\texttt{complete} function.

\spacingset{1}

\begin{Shaded}
\begin{Highlighting}[]
\DocumentationTok{\#\# Fit deterministic imputation model}
\NormalTok{imputation }\OtherTok{\textless{}{-}} \FunctionTok{mice}\NormalTok{(}\AttributeTok{m =} \DecValTok{1}\NormalTok{,}
                   \AttributeTok{method =} \StringTok{"norm.predict"}\NormalTok{,}
                   \AttributeTok{predictorMatrix =}\NormalTok{ predict\_mat\_noy,}
                   \AttributeTok{data =}\NormalTok{ data)}

\DocumentationTok{\#\# Extract imputations in a complete data set}
\NormalTok{data\_imp }\OtherTok{\textless{}{-}} \FunctionTok{complete}\NormalTok{(imputation)}
\end{Highlighting}
\end{Shaded}

\spacingset{1.9}

We can now use the filled-in data frame, \texttt{data\_imp} to fit the
outcome model specified in Equation~\ref{eq-outcome} as follows:

\spacingset{1}

\begin{Shaded}
\begin{Highlighting}[]
\NormalTok{outcome\_model }\OtherTok{\textless{}{-}} \FunctionTok{lm}\NormalTok{(y }\SpecialCharTok{\textasciitilde{}}\NormalTok{ x, }\AttributeTok{data =}\NormalTok{ data\_imp)}
\NormalTok{broom}\SpecialCharTok{::}\FunctionTok{tidy}\NormalTok{(outcome\_model) }
\end{Highlighting}
\end{Shaded}

\begin{longtable}[]{@{}lrrrr@{}}
\caption{Model estimates for the relationship between \(X\) and \(Y\)
after deterministic imputation}\tabularnewline
\toprule\noalign{}
Term & Estimate & Standard Error & Statistic & \(p\)-value \\
\midrule\noalign{}
\endfirsthead
\toprule\noalign{}
Term & Estimate & Standard Error & Statistic & \(p\)-value \\
\midrule\noalign{}
\endhead
\bottomrule\noalign{}
\endlastfoot
(Intercept) & -0.023 & 0.055 & -0.415 & 0.678 \\
x & 2.005 & 0.051 & 39.584 & 0.000 \\
\end{longtable}

\spacingset{1.9}

In the above example, we see the estimate for the coefficient for the
relationship between \(X\) and \(Y\) is approximately 2. The default
standard error reported in the summary output above is \emph{incorrect}
in the sense that it does not appropriately take into account fact that
we first performed an imputation step -- this standard error is
equivalent to taking the square root of the equation in
Equation~\ref{eq-var-beta}, what we refer to in
Section~\ref{sec-imp-var} as the \emph{model-based variance}. To
estimate the correct variance for this coefficient, what we refer to in
Section~\ref{sec-imp-var} as the \emph{true sampling variance}, we can
use a bootstrap. It is important when doing so we bootstrap the entire
process, including the imputation estimation. We begin by sampling the
rows of the original data frame, \texttt{data}, with replacement. We
then repeat the steps above to complete the imputation step as well as
fit the outcome model. We can then repeat this function \texttt{B}
times, in this case we will repeat it 1,000 times, and save the output
as \texttt{bootstrapped\_coefficients}. We can calculate the
bootstrapped standard error by taking the standard deviation of the
\texttt{bootstrapped\_coefficients} object.

\spacingset{1}

\begin{Shaded}
\begin{Highlighting}[]
\NormalTok{bootstrap }\OtherTok{\textless{}{-}} \ControlFlowTok{function}\NormalTok{() \{}
  \DocumentationTok{\#\# Sample the data rows with replacement}
\NormalTok{  data }\OtherTok{\textless{}{-}}\NormalTok{ data[}\FunctionTok{sample}\NormalTok{(}\FunctionTok{nrow}\NormalTok{(data), }\AttributeTok{replace =} \ConstantTok{TRUE}\NormalTok{), ]}

  \DocumentationTok{\#\# Create a matrix where we remove y from the imputation model}
\NormalTok{  predict\_mat }\OtherTok{\textless{}{-}} \FunctionTok{make.predictorMatrix}\NormalTok{(data)}
\NormalTok{  predict\_mat\_noy }\OtherTok{\textless{}{-}}\NormalTok{ predict\_mat}
\NormalTok{  predict\_mat\_noy[, }\StringTok{"y"}\NormalTok{] }\OtherTok{\textless{}{-}} \DecValTok{0}
  
  \DocumentationTok{\#\# Fit imputation model}
\NormalTok{  imputation }\OtherTok{\textless{}{-}} \FunctionTok{mice}\NormalTok{(}\AttributeTok{m =} \DecValTok{1}\NormalTok{,}
                     \AttributeTok{method =} \StringTok{"norm.predict"}\NormalTok{,}
                     \AttributeTok{predictorMatrix =}\NormalTok{ predict\_mat\_noy,}
                     \AttributeTok{data =}\NormalTok{ data,}
                     \AttributeTok{printFlag =} \ConstantTok{FALSE}\NormalTok{)}
  
  \DocumentationTok{\#\# Extract imputations in a complete data set}
\NormalTok{  data\_imp }\OtherTok{\textless{}{-}} \FunctionTok{complete}\NormalTok{(imputation)}
  
  \DocumentationTok{\#\# Fit outcome model}
\NormalTok{  outcome\_model }\OtherTok{\textless{}{-}} \FunctionTok{lm}\NormalTok{(y }\SpecialCharTok{\textasciitilde{}}\NormalTok{ x, }\AttributeTok{data =}\NormalTok{ data\_imp)}
  
  \DocumentationTok{\#\# Output coefficient}
  \FunctionTok{coef}\NormalTok{(outcome\_model)[}\StringTok{"x"}\NormalTok{]}
\NormalTok{\}}

\NormalTok{bootstrapped\_coefficients }\OtherTok{\textless{}{-}}\NormalTok{ purrr}\SpecialCharTok{::}\FunctionTok{map\_dbl}\NormalTok{(}\DecValTok{1}\SpecialCharTok{:}\DecValTok{1000}\NormalTok{, }\SpecialCharTok{\textasciitilde{}}\FunctionTok{bootstrap}\NormalTok{())}
\FunctionTok{sd}\NormalTok{(bootstrapped\_coefficients)}
\end{Highlighting}
\end{Shaded}

\spacingset{1.9}

The estimated standard error using the bootstrap method is 0.045, versus
the original, unadjusted model-based estimate 0.051 from above. The
bootstrap is estimating the correct variability, as it takes into
account the uncertainty in the imputation model step. If using this
methodology in practice, these are the standard errors that should be
reported, rather than the default model-based variability.

\hypertarget{stochastic-imputation}{%
\subsection{Stochastic Imputation}\label{stochastic-imputation}}

We begin by estimating \(X_{imp,stoc}\) as specified in
Equation~\ref{eq-imp-stoc}. By default, \texttt{mice} assumes we want to
include \texttt{y} in our imputation model, so we can leave the default
for stochastic imputation. We can complete single or multiple
imputation. Previous sections of this paper focus on single imputation,
however it is more efficient to use multiple imputation when
implementing stochastic imputation, therefore we will do so here. Now we
use the \texttt{mice} function with the following parameters:

\begin{itemize}
\tightlist
\item
  \texttt{m\ =\ 40} allows us to complete \emph{multiple} imputation
  with 40 imputations
\item
  \texttt{method\ =\ "norm"} allows us to apply \emph{stochastic}
  imputation
\end{itemize}

Since we completed multiple imputation, instead of extracting a single
data frame using the \texttt{complete} function, we will estimate the
outcome model in each imputed data set and pool the results using the
\texttt{pool} function, which applies Rubin's rules to estimate the
proper standard error.

\spacingset{1}

\begin{Shaded}
\begin{Highlighting}[]
\DocumentationTok{\#\# Fit imputation model}
\NormalTok{stoc\_imputation }\OtherTok{\textless{}{-}} \FunctionTok{mice}\NormalTok{(}\AttributeTok{m =} \DecValTok{40}\NormalTok{,}
                   \AttributeTok{method =} \StringTok{"norm"}\NormalTok{,}
                   \AttributeTok{data =}\NormalTok{ data)}

\DocumentationTok{\#\# Fit the outcome model}
\NormalTok{outcome\_model\_stoc }\OtherTok{\textless{}{-}} \FunctionTok{with}\NormalTok{(stoc\_imputation, }\FunctionTok{lm}\NormalTok{(y }\SpecialCharTok{\textasciitilde{}}\NormalTok{ x))}
\NormalTok{broom}\SpecialCharTok{::}\FunctionTok{tidy}\NormalTok{(}\FunctionTok{pool}\NormalTok{(outcome\_model\_stoc))}
\end{Highlighting}
\end{Shaded}

\begin{longtable}[]{@{}lrrrr@{}}
\caption{Model estimates for the relationship between \(X\) and \(Y\)
after stochastic imputation (M = 40)}\tabularnewline
\toprule\noalign{}
Term & Estimate & Standard Error & Statistic & \(p\)-value \\
\midrule\noalign{}
\endfirsthead
\toprule\noalign{}
Term & Estimate & Standard Error & Statistic & \(p\)-value \\
\midrule\noalign{}
\endhead
\bottomrule\noalign{}
\endlastfoot
(Intercept) & 0.017 & 0.041 & 0.402 & 0.688 \\
x & 2.019 & 0.034 & 58.769 & 0.000 \\
\end{longtable}

\spacingset{1.9}

In this simulation, the estimated standard error with multiple
stochastic imputation is 0.034.

When using the \texttt{mice} package to perform deterministic
imputation, it is important to use an appropriate method, such as the
bootstrap, to estimate the variability. When using the \texttt{mice}
package to perform stochastic imputation, it is likewise important to
use an appropriate method, such as Rubin's Rules via the \texttt{pool}
function, to estimate the variability. In both cases, the when the
imputation models are fit appropriately, as demonstrated here (i.e., the
deterministic imputation model does \emph{not} include the outcome and
the stochastic imputation model \emph{does} include the outcome), both
methods result in an unbiased estimate for the relationship between
\(X\) and \(Y\) (in this example, a coefficient of 2).

\hypertarget{discussion}{%
\section{Discussion}\label{discussion}}

This paper aims to bridge the gap between imputation in theory and in
practice by providing interpretable mathematical derivations of the
quantities required to calculate conditional statistics relying on
imputed covariates. Here we have investigated a simple case of missing
data in a covariate of interest. We demonstrated empirically, as well as
derived mathematically, the difference between deterministic and
stochastic imputation methods. These methods' impact on both the
variance of the imputed covariate as well as its relationship with a
continuous outcome is highlighted. Of note, in our running example even
though \(X\) is missing at random conditional on \(Z\), a complete case
analysis of \(Y\) on \(X_{obs}\) (without conditioning on \(Z\)) would
yield unbiased results, since \(Z\) is not a confounder (i.e., it has no
relationship with \(Y\) except through \(X\)). If \(Z\) were a
confounder, the same relationships shown would hold (i.e., the
deterministic imputation models would need \emph{not} include \(Y\) and
the stochastic imputation models would \emph{need} to include \(Y\)),
except the final outcome model would also need to adjust for \(Z\) in
order to yield an unbiased (now conditional) estimate of the effect of
\(X\) on \(Y\).

The recommendation to include the outcome in a stochastic imputation
model has been given (1,4,6,8). However, we have not seen a simple
mathematical explanation for why this is so. Additionally, the inclusion
of the outcome in the imputation model is often perceived as a
suggestion rather than a stipulation, but, as we demonstrate here, this
ought not just be \emph{recommended} but rather \emph{required} in order
to yield unbiased results with stochastic imputation. We have also seen
a disconnect in the recommendations dependent on the type of imputation
model fit (deterministic versus stochastic), as the deterministic models
\emph{should not} include the outcome, while the stochastic models
\emph{should}.

Multiple imputation is ubiquitous in biomedical research. While we focus
on conventionally missing data, imputation methods have also been used
to overcome covariate measurement error (25,26), and the same
considerations apply to these broader settings, as well. There have been
many documented examples of misspecified imputation models biasing
results (7,8,27,28). For example, as documented in (8), (27), and (7) a
widely-used clinical prediction model for cardiovascular disease, QRISK,
was initially run using stochastic imputation that excluded the outcome
from the imputation model. This exclusion resulted in estimates of the
coefficient for the ratio of cholesterol values, a known risk factor for
cardiovascular disease, to be biased towards the null (0.00 among both
women and men). It was pointed out in several rapid responses (e.g.,
(29)) that the imputation model was misspecified. When the authors
corrected the analysis to include the outcome in the imputation model,
they found updated coefficients of 3.22 among women and 3.30 among men
(both of which were statistically significant, consistent with prior
literature) (27). This example demonstrates the importance of
understanding which imputation methods are appropriate for biomedical
research.

In sum, deterministic imputation excluding the outcome yields unbiased
results when fitting a model after imputing covariates or variables of
interest. Corrections are needed to obtain valid variance estimates
after single imputation (14,23). If you are doing stochastic imputation,
you must include the outcome (and any other covariates in your final
analysis model) in your imputation model in order to get unbiased
coefficient estimates. If answering an inferential question, the Robins
and Wang asymptotic variance estimator can be used to obtain valid
variance estimates for single or multiple stochastic imputation (23), or
Rubin's Rules can be used along with multiple imputation (16). When
answering an inferential question, as we have demonstrated here, if the
imputation model is correctly specified both deterministic and
stochastic imputation will yield unbiased results. Multiple stochastic
imputation will outperform single deterministic imputation in terms of
efficiency, however it is conceivable to have situations with
sufficiently large sample sizes where the latter's ease of
implementation may outweigh its efficiency loss. Additionally, there are
clinical scenarios where the outcome may not be available at the time of
imputation, making deterministic imputation an appealing alternative
(30). A recent simulation study supports the results we show here,
specifically highlighting that in the context of clinical prediction
models, deterministic imputation performs comparably to stochastic
imputation (6). We hope the mathematical demonstrations in this paper
serve to strengthen this result and further elucidate the reasons behind
current recommendations for practitioners.

\hypertarget{references}{%
\section{References}\label{references}}

\hypertarget{refs}{}
\begin{CSLReferences}{0}{0}
\leavevmode\vadjust pre{\hypertarget{ref-moons2006using}{}}%
\CSLLeftMargin{1. }%
\CSLRightInline{Moons KG, Donders RA, Stijnen T, Harrell Jr FE. Using
the outcome for imputation of missing predictor values was preferred.
Journal of Clinical Epidemiology. 2006;59(10):1092--101. }

\leavevmode\vadjust pre{\hypertarget{ref-lee2012recovery}{}}%
\CSLLeftMargin{2. }%
\CSLRightInline{Lee KJ, Carlin JB. Recovery of information from multiple
imputation: A simulation study. Emerging Themes in Epidemiology.
2012;9:1--10. }

\leavevmode\vadjust pre{\hypertarget{ref-jochen2013multiple}{}}%
\CSLLeftMargin{3. }%
\CSLRightInline{Jochen H, Max H, Tamara B, Wilfried L. Multiple
imputation of missing data: A simulation study on a binary response.
Open Journal of Statistics. 2013;2013. }

\leavevmode\vadjust pre{\hypertarget{ref-kontopantelis2017outcome}{}}%
\CSLLeftMargin{4. }%
\CSLRightInline{Kontopantelis E, White IR, Sperrin M, Buchan I.
Outcome-sensitive multiple imputation: A simulation study. BMC Medical
Research Methodology. 2017;17(1):1--13. }

\leavevmode\vadjust pre{\hypertarget{ref-sperrin2020multiple}{}}%
\CSLLeftMargin{5. }%
\CSLRightInline{Sperrin M, Martin GP. Multiple imputation with missing
indicators as proxies for unmeasured variables: Simulation study. BMC
Medical Research Methodology. 2020;20(1):1--11. }

\leavevmode\vadjust pre{\hypertarget{ref-sisk2023imputation}{}}%
\CSLLeftMargin{6. }%
\CSLRightInline{Sisk R, Sperrin M, Peek N, Smeden M van, Martin GP.
Imputation and missing indicators for handling missing data in the
development and deployment of clinical prediction models: A simulation
study. Statistical Methods in Medical Research. 2023;09622802231165001.
}

\leavevmode\vadjust pre{\hypertarget{ref-sterne2009multiple}{}}%
\CSLLeftMargin{7. }%
\CSLRightInline{Sterne JA, White IR, Carlin JB, Spratt M, Royston P,
Kenward MG, et al. Multiple imputation for missing data in
epidemiological and clinical research: Potential and pitfalls. BMJ.
2009;338. }

\leavevmode\vadjust pre{\hypertarget{ref-bartlett2011multiple}{}}%
\CSLLeftMargin{8. }%
\CSLRightInline{Bartlett JW, Frost C, Carpenter JR. Multiple imputation
models should incorporate the outcome in the model of interest. Brain.
2011;134(11):e189--9. }

\leavevmode\vadjust pre{\hypertarget{ref-donders2006gentle}{}}%
\CSLLeftMargin{9. }%
\CSLRightInline{Donders ART, Van Der Heijden GJ, Stijnen T, Moons KG. A
gentle introduction to imputation of missing values. Journal of Clinical
Epidemiology. 2006;59(10):1087--91. }

\leavevmode\vadjust pre{\hypertarget{ref-greenland1995critical}{}}%
\CSLLeftMargin{10. }%
\CSLRightInline{Greenland S, Finkle WD. A critical look at methods for
handling missing covariates in epidemiologic regression analyses.
American journal of epidemiology. 1995;142(12):1255--64. }

\leavevmode\vadjust pre{\hypertarget{ref-little1992regression}{}}%
\CSLLeftMargin{11. }%
\CSLRightInline{Little RJ. Regression with missing x's: A review.
Journal of the American Statistical Association. 1992;87(420):1227--37.
}

\leavevmode\vadjust pre{\hypertarget{ref-mice}{}}%
\CSLLeftMargin{12. }%
\CSLRightInline{van Buuren S, Groothuis-Oudshoorn K. {mice}:
Multivariate imputation by chained equations in r. Journal of
Statistical Software. 2011;45(3):1--67. }

\leavevmode\vadjust pre{\hypertarget{ref-moreno2018canonical}{}}%
\CSLLeftMargin{13. }%
\CSLRightInline{Moreno-Betancur M, Lee KJ, Leacy FP, White IR, Simpson
JA, Carlin JB. Canonical causal diagrams to guide the treatment of
missing data in epidemiologic studies. American Journal of Epidemiology.
2018;187(12):2705--15. }

\leavevmode\vadjust pre{\hypertarget{ref-schafer2000inference}{}}%
\CSLLeftMargin{14. }%
\CSLRightInline{Schafer JL, Schenker N. Inference with imputed
conditional means. Journal of the American Statistical Association.
2000;95(449):144--54. }

\leavevmode\vadjust pre{\hypertarget{ref-shao2013estimation}{}}%
\CSLLeftMargin{15. }%
\CSLRightInline{Shao J. Estimation and imputation in linear regression
with missing values in both response and covariate. Statistics and its
Interface. 2013;6(3):361--8. }

\leavevmode\vadjust pre{\hypertarget{ref-rubin2004multiple}{}}%
\CSLLeftMargin{16. }%
\CSLRightInline{Rubin DB. Multiple imputation for nonresponse in
surveys. Vol. 81. John Wiley \& Sons; 2004. }

\leavevmode\vadjust pre{\hypertarget{ref-van2018flexible}{}}%
\CSLLeftMargin{17. }%
\CSLRightInline{Van Buuren S. Flexible imputation of missing data. CRC
press; 2018. }

\leavevmode\vadjust pre{\hypertarget{ref-daniels2014fully}{}}%
\CSLLeftMargin{18. }%
\CSLRightInline{Daniels M, Wang C, Marcus B. Fully bayesian inference
under ignorable missingness in the presence of auxiliary covariates.
Biometrics. 2014;70(1):62--72. }

\leavevmode\vadjust pre{\hypertarget{ref-bartlett2020bootstrap}{}}%
\CSLLeftMargin{19. }%
\CSLRightInline{Bartlett JW, Hughes RA. Bootstrap inference for multiple
imputation under uncongeniality and misspecification. Statistical
Methods in Medical Research. 2020;29(12):3533--46. }

\leavevmode\vadjust pre{\hypertarget{ref-von20098}{}}%
\CSLLeftMargin{20. }%
\CSLRightInline{Von Hippel PT. 8. How to impute interactions, squares,
and other transformed variables. Sociological Methodology.
2009;39(1):265--91. }

\leavevmode\vadjust pre{\hypertarget{ref-white2011multiple}{}}%
\CSLLeftMargin{21. }%
\CSLRightInline{White IR, Royston P, Wood AM. Multiple imputation using
chained equations: Issues and guidance for practice. Statistics in
Medicine. 2011;30(4):377--99. }

\leavevmode\vadjust pre{\hypertarget{ref-von2013should}{}}%
\CSLLeftMargin{22. }%
\CSLRightInline{Hippel PT von. Should a normal imputation model be
modified to impute skewed variables? Sociological Methods \& Research.
2013;42(1):105--38. }

\leavevmode\vadjust pre{\hypertarget{ref-robins2000inference}{}}%
\CSLLeftMargin{23. }%
\CSLRightInline{Robins JM, Wang N. Inference for imputation estimators.
Biometrika. 2000;87(1):113--24. }

\leavevmode\vadjust pre{\hypertarget{ref-little2019statistical}{}}%
\CSLLeftMargin{24. }%
\CSLRightInline{Little RJ, Rubin DB. Statistical analysis with missing
data. Vol. 793. John Wiley \& Sons; 2019. }

\leavevmode\vadjust pre{\hypertarget{ref-cole2006multiple}{}}%
\CSLLeftMargin{25. }%
\CSLRightInline{Cole SR, Chu H, Greenland S. Multiple-imputation for
measurement-error correction. International Journal of Epidemiology.
2006;35(4):1074--81. }

\leavevmode\vadjust pre{\hypertarget{ref-shepherd2012using}{}}%
\CSLLeftMargin{26. }%
\CSLRightInline{Shepherd BE, Shaw PA, Dodd LE. Using audit information
to adjust parameter estimates for data errors in clinical trials.
Clinical Trials. 2012;9(6):721--9. }

\leavevmode\vadjust pre{\hypertarget{ref-hippisley2007derivation}{}}%
\CSLLeftMargin{27. }%
\CSLRightInline{Hippisley-Cox J, Coupland C, Vinogradova Y, Robson J,
May M, Brindle P. Derivation and validation of QRISK, a new
cardiovascular disease risk score for the united kingdom: Prospective
open cohort study. BMJ. 2007;335(7611):136. }

\leavevmode\vadjust pre{\hypertarget{ref-hippisley2007qrisk}{}}%
\CSLLeftMargin{28. }%
\CSLRightInline{Hippisley-Cox J, Coupland C, Vinogradova Y, Robson J,
Brindle P. QRISK cardiovascular disease risk prediction algorithm {œ}
comparison of the revised and the original analyses technical
supplement. qresearchorg. 2007; }

\leavevmode\vadjust pre{\hypertarget{ref-response}{}}%
\CSLLeftMargin{29. }%
\CSLRightInline{Carlin J. Rapid response: Multiple imputation needs to
be used with care and reported in detail. BMJ. 2007;335(7611):136. }

\leavevmode\vadjust pre{\hypertarget{ref-fletcher2020missing}{}}%
\CSLLeftMargin{30. }%
\CSLRightInline{Fletcher Mercaldo S, Blume JD. Missing data and
prediction: The pattern submodel. Biostatistics. 2020;21(2):236--52. }

\leavevmode\vadjust pre{\hypertarget{ref-tidyverse}{}}%
\CSLLeftMargin{31. }%
\CSLRightInline{Wickham H, Averick M, Bryan J, Chang W, D'Agostino
McGowan L, François R, et al. Welcome to the {tidyverse}. Journal of
Open Source Software. 2019;4(43):1686. }

\leavevmode\vadjust pre{\hypertarget{ref-R}{}}%
\CSLLeftMargin{32. }%
\CSLRightInline{R Core Team. R: A language and environment for
statistical computing {[}Internet{]}. Vienna, Austria: R Foundation for
Statistical Computing; 2022. Available from:
\url{https://www.R-project.org/}}

\end{CSLReferences}

\hypertarget{appendix-a---derivation-explanations}{%
\section*{Appendix A - Derivation
Explanations}\label{appendix-a---derivation-explanations}}
\addcontentsline{toc}{section}{Appendix A - Derivation Explanations}

\spacingset{1}

The following are explanations for derivations seen in Section 2. For
our motivating example, the relationship between \(X\) and \(Z\) is as
follows:

\[
X = Z + \varepsilon_X \textrm{, where } Z\sim Bern(0.5) \textrm{ and }\varepsilon_X\sim N(0,1).
\]

Let \(R_{X}\) be the a missingness indicator for the observed \(x\) were
we define \(R_X\) such that

\[
\Pr(R_X = 1|X,Z) = \Pr(R_X = 1|Z) = 0.25(1 - Z) + 0.50(Z).
\]

We will begin with the intuition behind the formula for
\(\textrm{Var}(X|R_X=0)\) in Equation~\ref{eq-var}. This quantity is the
variance of the observed (i.e., non-missing) values of \(X\). We know
that the conditional probabilities of observing \(X\) are
\(\Pr(R_X=0|Z=1)=0.5\) and \(\Pr(R_X=0|Z=0)=0.75\), with the marginal
probability \(\Pr(R_X=0)=0.625\). Since the marginal probability of
\(Z\) is \(\Pr(Z=1)=0.5\), we know
\(\Pr(Z=1|R_X=0) = \Pr(R_X=0|Z=1)\Pr(Z=1) / \Pr(R_X=0) = (0.5)(0.5)/(0.625) = 0.4\).
Likewise, \(\Pr(Z=0|R_X=0)=(0.75)(0.5)/(0.625) =0.6\). Therefore, by the
law of total variance

\[
\begin{aligned}
\textrm{Var}(X|R_X=0) =& \textrm{E}\{\textrm{Var}(X|Z,R_X=0)|R_X=0\}+\textrm{Var}\{\textrm{E}(X|Z,R_X=0)|R_X=0\}\\
=&\textrm{E}\{\textrm{Var}(X|Z,R_X=0)|R_X=0\}+\textrm{Var}(Z|R_X=0)\\
=&\textrm{Var}(X|Z=1,R_X=0)\Pr(Z=1|R_X=0)\\
&+\textrm{Var}(X|Z=0, R_X=0)\Pr(Z=0|R_X=0)\\
&+\Pr(Z=1|R_X=0)\Pr(Z=0|R_X=0)\\
=&1(0.6) + 1(0.4) + 0.6(0.4)\\
=&1.24.
\end{aligned}
\]

Next, we will show why
\(\textrm{Var}(\widehat{X}_{det}|R_X=1)= 0.\overline{22}\). Since
\(\Pr(R_X=1|Z=1)=0.5\) and \(\Pr(R_X=1|Z=0)=0.25\), we know
\(\Pr(Z=1|R_X=1)=(0.5)(0.5)/(0.375) = 0.\overline{66}\) and
\(\Pr(Z=0|R_X=1)=(0.25)(0.5)/(0.375) = 0.\overline{33}\). Therefore, the
expected variance is as follows:

\[
\begin{aligned}
\textrm{Var}(\widehat{X}_{det}|R_X=1)=&\textrm{Var}(\widehat\alpha_0+\widehat\alpha_1Z|R_X=1) \\
=&\alpha_1^2\textrm{Var}(Z|R_X=1)\\
=&(1)(0.\overline{66})(0.\overline{33})\\
=&0.\overline{22}.
\end{aligned}
\]

Note that we are referring to the quantity above as the expected
variance, by which we mean the variance evaluated at the expected value
of the imputed value (where \(E(\widehat\alpha_1) = \alpha_1\)). The
expected value of the non-missing \(X\), \(\textrm{E}(X|R_X=0)\), is
0.4. Because \(X\) has a mean of 1 when \(Z=1\) and a mean of 0 when
\(Z=0\), and we know that \(\Pr(Z=1|R_X=0)=0.4\), we therefore know the
following:

\[
\begin{aligned}
\textrm{E}(X|R_X=0)=&\textrm{E}(X|Z=0,R_X=0)\Pr(Z=0|R_X=0) \\
&+\textrm{E}(X|Z=1,R_X=0)\Pr(Z=1|R_X=0)\\
=&0(0.6) + 1(0.4)\\
=&0.4.
\end{aligned}
\] We can show that
\(\textrm{E}(\widehat{X}_{det}|R_X=1)=0.\overline{66}\) as follows:

\[
\begin{aligned}
\textrm{E}(\widehat{X}_{det}|R_X=1)=&\textrm{E}(\widehat\alpha_0 + \widehat\alpha_1Z|R_X=1)\\
=&\alpha_0 + \alpha_1\textrm{E}(Z|R_X=1)\\
=&0+1(0.\overline{66})\\
=&0.\overline{66}.
\end{aligned}
\]

\hypertarget{sec-imp-sub-derivations}{%
\subsubsection*{\texorpdfstring{Section~\ref{sec-imp-sub}
Derivations}{Section~ Derivations}}\label{sec-imp-sub-derivations}}
\addcontentsline{toc}{subsubsection}{Section~\ref{sec-imp-sub}
Derivations}

Moving on to Section~\ref{sec-imp-sub}, we can show that
\(\textrm{Cov}(X, Y|R_X=0)\) is as follows:

\[
\begin{aligned}
\textrm{Cov}(X, Y|R_X=0) =& \textrm{Cov}(X, 2X+\varepsilon_Y|R_X=0)\\
=&2\textrm{Var}(X|R_X=0)\\
=&2(1.24)  =2.48.
\end{aligned}
\]

This quantity is the covariance between the non-missing \(X\) and the
corresponding values of \(Y\) for which \(X\) was not missing. We can
also show that the expected \(\textrm{Cov}(\widehat{X}_{det}, Y|R_X=1)\)
is as follows:

\[
\begin{aligned}
\textrm{Cov}(\widehat{X}_{det}, Y|R_X=1)=&\textrm{Cov}(\widehat\alpha_0+\widehat\alpha_1Z, Y|R_X=1)\\
=&\alpha_1\textrm{Cov}(Z, Y|R_X=1)\\
=&\textrm{Cov}(Z, 2X+\varepsilon_Y|R_X=1)\\
=&2\textrm{Cov}(Z, X|R_X=1)\\
=&2\textrm{Var}(Z|R_X=1)\\
=&2(0.\overline{22})\\
=&0.\overline{44}.
\end{aligned}
\] The mean of \(Y\) among those with missing values of \(X\) is:

\[
\begin{aligned}
\textrm{E}(Y|R_X=0) &= \textrm{E}(2X+\varepsilon_Y|R_X=0)\\
&=2\textrm{E}(X|R_X=0)\\
&=2(0.4) = 0.8.
\end{aligned}
\]

The mean of \(Y\) among those with non-missing values of \(X\) is:

\[
\begin{aligned}
\textrm{E}(Y|R_X=1) &= \textrm{E}(2X+\varepsilon_Y|R_X=1)\\
&=2\textrm{E}(X|R_X=1)\\
&=2(0.\overline{66}) = 1.\overline{33}.
\end{aligned}
\]

\hypertarget{sec-imp-var-derivations}{%
\subsubsection*{\texorpdfstring{Section~\ref{sec-imp-var}
Derivations}{Section~ Derivations}}\label{sec-imp-var-derivations}}
\addcontentsline{toc}{subsubsection}{Section~\ref{sec-imp-var}
Derivations}

Based on Equation~\ref{eq-y}, we know that the variance of \(Y\) is:

\[
\begin{aligned}
\textrm{Var}(Y) &= \textrm{Var}(2X + \varepsilon_Y)\\
&=4\textrm{Var}(X) + \textrm{Var}(\varepsilon_Y)\\
&=4(1.25) + 1\\
&= 6.
\end{aligned}
\]

We can begin by calculating the expected variance of
\(\widehat\beta_{1, Y\sim X}\) had \(X\) been fully observed:

\[
\begin{aligned}
{\widehat{\textrm{Var}}}\left(\widehat\beta_{1,Y\sim X}\right) &= \left\{\frac{\widehat{\textrm{Var}}(Y)}{\widehat{\textrm{Var}}(X)}-\frac{\widehat{\textrm{Cov}}(X,Y)^2}{\widehat{\textrm{Var}}(X)^2}\right\}\left(\frac{1}{n-2}\right)\\
&=\left(\frac{6}{1.25}-\frac{2.5^2}{1.25^2}\right)\left(\frac{1}{n-2}\right)\\
&= (4.8-4)\left(\frac{1}{n-2}\right)\\
&= \frac{0.8}{n-2}.\\
\end{aligned}
\]

Using this quantity along with Equation~\ref{eq-var-beta}, we can show
that the expected variance of \(\widehat\beta_{1,Y\sim X_{imp,det}}\)
assuming \(X_{imp,det}\) is fixed is:

\[
\begin{aligned}
{\widehat{\textrm{Var}}}\left(\widehat\beta_{1,Y\sim X_{imp,det}}|X_{imp,det}\right) &= \left\{\frac{\widehat{\textrm{Var}}(Y)}{\widehat{\textrm{Var}}(X_{imp,det})}-\frac{\widehat{\textrm{Cov}}(X_{imp,det},Y)^2}{\widehat{\textrm{Var}}(X_{imp,det})^2}\right\}\left(\frac{1}{n-2}\right)\\
&=\left(\frac{6}{0.875}-\frac{1.75^2}{0.875^2}\right)\left(\frac{1}{n-2}\right)\\
&= \left(\frac{6}{0.875}-4\right)\left(\frac{1}{n-2}\right)\\
&= \frac{2.857143}{n-2}.\\
\end{aligned}
\] Finally, to calculate the complete case coefficient's expected
variance, we need the variance of \(Y\) conditional on \(R_X=0\):

\[
\begin{aligned}
\textrm{Var}(Y|R_X=0) &= \textrm{Var}(2X + \varepsilon_Y|R_X=0)\\
&=4\textrm{Var}(X|R_X=0) + \textrm{Var}(\varepsilon_Y)\\
&=4(1.24) + 1\\
&=5.96.
\end{aligned}
\]

We can use this quantity to show that the expected variance of the
complete case estimator, \(\widehat\beta_{1,Y\sim X|R_X=0}\), is as
follows:

\[
\begin{aligned}
\widehat{\textrm{Var}}\left(\widehat\beta_{1,Y\sim X|R_X=0}\right) &= \left\{\frac{\widehat{\textrm{Var}}(Y|R_X=0)}{\widehat{\textrm{Var}}(X|R_X=0)}-\frac{\widehat{\textrm{Cov}}(X,Y|R_X=0)^2}{\widehat{\textrm{Var}}(X|R_X=0)^2}\right\}\left(\frac{1}{n_{obs}-2}\right)\\
&=\left(\frac{5.96}{1.24}-\frac{2.48^2}{1.24^2}\right)\left(\frac{1}{n_{obs}-2}\right)\\
&= \left(\frac{5.96}{1.24}-4\right)\left(\frac{1}{n_{obs}-2}\right)\\
&= \frac{0.806452}{n_{obs}-2}\\
&=\frac{0.806452}{n\Pr(R_X=0)-2}\\
&=\frac{0.806452}{n(0.625)-2}.
\end{aligned}
\]

For example, if \(n = 102\), on average
\(\widehat{\textrm{Var}}\left(\widehat\beta_{1,Y\sim X}\right) = 0.008\),
\(\widehat{\textrm{Var}}\left(\widehat\beta_{1,Y\sim X_{imp,det}} | X_{imp,det}\right) \approx 0.029\),
and
\(\widehat{\textrm{Var}}\left(\widehat\beta_{1,Y\sim X|R_X=0}\right) \approx 0.013\).
We can compare these values to
\(\widehat{\textrm{Var}}\left(\widehat\beta_{1,Y\sim X_{imp,det}}\right)\)
(that is, the variance where we do \emph{not} assume the imputed values
are fixed), which in this example is approximately \(0.02\), as
demonstrated in the simulation below.

\begin{Shaded}
\begin{Highlighting}[]
\NormalTok{n }\OtherTok{\textless{}{-}} \DecValTok{102}
\NormalTok{sim\_determ }\OtherTok{\textless{}{-}} \ControlFlowTok{function}\NormalTok{() \{}
\NormalTok{  data }\OtherTok{\textless{}{-}} \FunctionTok{tibble}\NormalTok{(}
    \AttributeTok{z =} \FunctionTok{rbinom}\NormalTok{(n, }\DecValTok{1}\NormalTok{, }\AttributeTok{p =} \FloatTok{0.5}\NormalTok{),}
    \AttributeTok{x =}\NormalTok{ z }\SpecialCharTok{+} \FunctionTok{rnorm}\NormalTok{(n),}
    \AttributeTok{y =} \DecValTok{2} \SpecialCharTok{*}\NormalTok{ x }\SpecialCharTok{+} \FunctionTok{rnorm}\NormalTok{(n),}
    \AttributeTok{x\_miss =} \FunctionTok{ifelse}\NormalTok{(z, }\FunctionTok{rbinom}\NormalTok{(n, }\DecValTok{1}\NormalTok{, }\FloatTok{0.5}\NormalTok{), }\FunctionTok{rbinom}\NormalTok{(n, }\DecValTok{1}\NormalTok{, }\FloatTok{0.25}\NormalTok{)),}
    \AttributeTok{x\_obs =} \FunctionTok{ifelse}\NormalTok{(x\_miss, }\ConstantTok{NA}\NormalTok{, x)}
\NormalTok{  )}
\NormalTok{  imp\_fit }\OtherTok{\textless{}{-}} \FunctionTok{lm}\NormalTok{(x\_obs }\SpecialCharTok{\textasciitilde{}}\NormalTok{ z, }\AttributeTok{data =}\NormalTok{ data)}
\NormalTok{  data }\OtherTok{\textless{}{-}}\NormalTok{ data }\SpecialCharTok{|\textgreater{}}
    \FunctionTok{mutate}\NormalTok{(}\AttributeTok{x\_imp =} \FunctionTok{ifelse}\NormalTok{(x\_miss, }\FunctionTok{predict}\NormalTok{(imp\_fit, }\AttributeTok{newdata =}\NormalTok{ data), x\_obs))}
  \FunctionTok{coef}\NormalTok{(}\FunctionTok{lm}\NormalTok{(y }\SpecialCharTok{\textasciitilde{}}\NormalTok{ x\_imp, }\AttributeTok{data =}\NormalTok{ data))[}\DecValTok{2}\NormalTok{]}
\NormalTok{\}}
\NormalTok{o }\OtherTok{\textless{}{-}}\NormalTok{ purrr}\SpecialCharTok{::}\FunctionTok{map\_dbl}\NormalTok{(}\DecValTok{1}\SpecialCharTok{:}\DecValTok{1000000}\NormalTok{, }\SpecialCharTok{\textasciitilde{}}\FunctionTok{sim\_determ}\NormalTok{())}
\FunctionTok{var}\NormalTok{(o)}
\end{Highlighting}
\end{Shaded}

\begin{verbatim}
[1] 0.02074008
\end{verbatim}

\hypertarget{sec-imp-y-derivations}{%
\subsubsection*{\texorpdfstring{Section~\ref{sec-imp-y}
Derivations}{Section~ Derivations}}\label{sec-imp-y-derivations}}
\addcontentsline{toc}{subsubsection}{Section~\ref{sec-imp-y}
Derivations}

Let's begin by deriving \(\textrm{Var}(X|R_X=1)\):

\[
\begin{aligned}
\textrm{Var}(X|R_X=1) = &\textrm{Var}(X|Z=1,R_X=1)\Pr(Z=1|R_X=1)\\
&+\textrm{Var}(X|Z=0, R_X=1)\Pr(Z=0|R_X=1)\\
&+\Pr(Z=1|R_X=1)\Pr(Z=0|R_X=1)\\
=&1(0.\overline{66}) + 1(0.\overline{33}) + 0.\overline{66}(0.\overline{33})\\
=&1.\overline{22}.
\end{aligned}
\]

Using this quantity, we can show that:

\[
\begin{aligned}
\textrm{Var}(\widehat{X}_{det|y}|R_X=1)=&\textrm{Var}(\widehat\gamma_0+\widehat\gamma_1Z+\widehat\gamma_2Y|R_X=1)\\
=&\gamma_1^2\textrm{Var}(Z|R_X=1)+\gamma_2^2\textrm{Var}(Y|R_X=1)+2\gamma_1\gamma_2\textrm{Cov}(Z,Y|R_X=1)\\
=&\gamma_1^2\textrm{Var}(Z|R_X=1)+\gamma_2^2\textrm{Var}(2X+\varepsilon_Y|R_X=1)+2\gamma_1\gamma_2\textrm{Cov}(Z,2X+\varepsilon_Y|R_X=1)\\
=&\gamma_1^2\textrm{Var}(Z|R_X=1)+\gamma_2^2(4)\textrm{Var}(X|R_X=1)+\gamma_2^2\textrm{Var}(\varepsilon_Y|R_X=1)\\
&+4\gamma_1\gamma_2\textrm{Cov}(Z,X|R_X=1)\\
=&\gamma_1^2\textrm{Var}(Z|R_X=1)+\gamma_2^2(4)\textrm{Var}(X|R_X=1)+\gamma_2^2\textrm{Var}(\varepsilon_Y|R_X=1)\\
&+4\gamma_1\gamma_2\textrm{Cov}(Z,Z+\varepsilon_X|R_X=1)\\
=&\gamma_1^2\textrm{Var}(Z|R_X=1)+\gamma_2^2(4)\textrm{Var}(X|R_X=1)+\gamma_2^2\textrm{Var}(\varepsilon_Y|R_X=1)\\
&+4\gamma_1\gamma_2\textrm{Var}(Z|R_X=1)\\
=&0.2^2(0.\overline{22})+0.4^2(4)(1.\overline{22})+0.4^2(1)+4(0.2)(0.4)(0.\overline{22})\\
=&1.0\overline{22}.
\end{aligned}
\]

Plugging \(\textrm{Var}(\widehat{X}_{det|y}|R_X=1)\) into
Equation~\ref{eq-var}, we get:

\[
\begin{aligned}
\textrm{Var}(X_{imp,det}) =1.24(0.625) + 1.0\overline{22}(0.375)+(0.4 - 0.\overline{66})^2(0.625)(0.375) = 1.175.
\end{aligned}
\]

We can also show that, evaluated at the expected values of the
imputation model,
\(\textrm{Cov}(\widehat{X}_{det|y},Y|R_X=1) = 2.\overline{44}\),
calculated as

\[
\begin{aligned}
\textrm{Cov}(\widehat{X}_{det|y},Y|R_X=1)&=\textrm{Cov}(\widehat\gamma_0+\widehat\gamma_1Z+\widehat\gamma_2Y,Y|R_X=1)\\
&=\gamma_1\textrm{Cov}(Z,Y|R_X=1)+\gamma_2\textrm{Var}(Y|R_X=1)\\
&=\gamma_1\textrm{Cov}(Z,Y|R_X=1)+\gamma_2\textrm{Var}(Y|R_X=1)\\
&=\gamma_1\textrm{Cov}(Z,2X+\varepsilon_Y|R_X=1)+\gamma_2\textrm{Var}(2X+\varepsilon_Y|R_X=1)\\
&=\gamma_1(2)\textrm{Cov}(Z,X|R_X=1)+\gamma_2(4)\textrm{Var}(X|R_X=1) +\gamma_2\textrm{Var}(\varepsilon|R_X=1)\\
&=\gamma_1(2)\textrm{Var}(Z|R_X=1)+\gamma_2(4)\textrm{Var}(X|R_X=1) +\gamma_2\textrm{Var}(\varepsilon|R_X=1)\\
&=0.2(2)(0.\overline{22}) + 0.4(4)(1.\overline{22})+0.4(1)\\
&=2.\overline{44}.
\end{aligned}
\]

Therefore,

\[
\begin{aligned}
\textrm{Cov}(X_{imp,det|y}, Y) &=2.48(0.625) + 2.\overline{44}(0.375) + 0.8(0.4)(0.625) + 1.\overline{33}(0.\overline{66})(0.375) - 0.5 \\&= 2.5.
\end{aligned}
\]

\hypertarget{deterministic-imputation-covariancevariance-ratio}{%
\subsection*{Deterministic imputation covariance/variance
ratio}\label{deterministic-imputation-covariancevariance-ratio}}
\addcontentsline{toc}{subsection}{Deterministic imputation
covariance/variance ratio}

We can write
\(X_{imp, det}=(1-R_X)X + R_X\widehat{X}_{det}=X - R_X\widehat\varepsilon_X\),
where \(\widehat\varepsilon_X\) is the estimated residual from the model
predicting \(X\) as in Equation~\ref{eq-1}. We can also write the
\(\textrm{Var}(\widehat\varepsilon_X)\) with respect to the \(R^2\) from
the model fit as in Equation~\ref{eq-1} as follows.

\[
\begin{aligned}
\textrm{Var}(\widehat\varepsilon_X|R_X=1) =& \textrm{Var}(X-\widehat{X}_{det}|R_X=1)\\
=& \textrm{Var}(X-(\widehat\alpha_0+\widehat\alpha_1Z)|R_X=1)\\
=&\textrm{Var}(X-\widehat\alpha_1Z|R_X=1)\\
=&\textrm{Var}(X|R_X=1)+\widehat\alpha_1^2\textrm{Var}(Z|R_X=1)-2\widehat\alpha_1\textrm{Cov}(X, Z|R_X=1)\\
=&\textrm{Var}(X|R_X=1)+R^2\frac{\textrm{Var}(X|R_X=0)}{\textrm{Var}(Z|R_X=0)}\textrm{Var}(Z|R_X=1)\\
&-2\widehat\alpha_1\textrm{Cov}(X, Z|R_Z=1)\\
=&\textrm{Var}(X|R_X=1)+R^2\frac{\textrm{Var}(X|R_X=0)}{\textrm{Var}(Z|R_X=0)}\textrm{Var}(Z|R_X=1)\\
&-2\frac{\textrm{Cov}(X, Z|R_X=1)\textrm{Cov}(X, Z|R_X=0)}{\textrm{Var}(Z|R_X=0)}\\
=&\textrm{Var}(X|R_X=1)+R^2\frac{\textrm{Var}(X|R_X=0)}{\textrm{Var}(Z|R_X=0)}\textrm{Var}(Z|R_X=1)\\
&-2\frac{R^2\textrm{Var}(X|R_X=0)\textrm{Cov}(X, Z|R_X=1)}{\textrm{Cov}(X, Z|R_X=0)}\\
=&\textrm{Var}(X|R_X=1)\\
&+R^2\textrm{Var}(X|R_X=0)\left(\frac{\textrm{Var}(Z|R_X=1)}{\textrm{Var}(Z|R_X=0)}-2\frac{\textrm{Cov}(X, Z|R_X=1)}{\textrm{Cov}(X, Z|R_X=0)}\right)\\
=&\textrm{Var}(X|R_X=1)-R^2\textrm{Var}(X|R_X=0)\left(\frac{\textrm{Var}(Z|R_X=1)}{\textrm{Var}(Z|R_X=0)}\right)\\
\end{aligned}
\]

Note that:

\[
\begin{aligned}
\frac{\textrm{Var}(Z|R_X=1)}{\textrm{Var}(Z|R_X=0)}&=\frac{\textrm{Cov}(X,Z|R_X=1)}{\textrm{Cov}(X,Z|R_X=0)}\\
&=\frac{\textrm{Cov}(\alpha_0+\alpha_1Z+\varepsilon_X,Z|R_X=1)}{\textrm{Cov}(\alpha_0+\alpha_1Z+\varepsilon_X,Z|R_X=0)}\\
&=\frac{\alpha_1\textrm{Cov}(Z,Z|R_X=1)+\textrm{Cov}(\varepsilon_X,Z|R_X=1)}{\alpha_1\textrm{Cov}(Z, Z|R_X=0)+\textrm{Cov}(\varepsilon_X,Z|R_X=0)}\\
&=\frac{\alpha_1\textrm{Var}(Z|R_X=1)}{\alpha_1\textrm{Var}(Z|R_X=0)}\\
&=\frac{\textrm{Var}(Z|R_X=1)}{\textrm{Var}(Z|R_X=0)}\\
\end{aligned}
\]

We can then write \(\textrm{Var}(X_{imp,det})\) as a function of
\(\textrm{Var}(X)\) as follows:

\[
\begin{aligned}
\textrm{Var}(X_{imp, det}) =& \textrm{Var}(X - R_X\widehat\varepsilon_X)\\
=&\textrm{Var}(X) + \textrm{Var}(R_X\widehat\varepsilon_X) - 2\textrm{Cov}(X, R_X\widehat\varepsilon_X)\\
=&\textrm{Var}(X) + \underbrace{\textrm{Var}\{\textrm{E}(-R_X\widehat\varepsilon_X|R_X)\}}_{=0} + \textrm{E}\{\textrm{Var}(-R_X\widehat\varepsilon_X|R_X)\} -2\textrm{Cov}(X, R_X\widehat\varepsilon_X)\\
=&\textrm{Var}(X) + \textrm{Var}(X|R_X=1)\Pr(R_X=1)\\
&-R^2\textrm{Var}(X|R_X=0)\frac{\textrm{Var}(Z|R_X=1)}{\textrm{Var}(Z|R_X=0)}\Pr(R_X=1) \\
&-2\textrm{Cov}(X, R_X(X-\widehat{X}_{det}))\\
=&\textrm{Var}(X) + \textrm{Var}(X|R_X=1)\Pr(R_X=1)\\
&-R^2\textrm{Var}(X|R_X=0)\frac{\textrm{Var}(Z|R_X=1)}{\textrm{Var}(Z|R_X=0)}\Pr(R_X=1) \\
&-2\{\textrm{Cov}(X,X|R_X=1)\Pr(R_X=1) - \textrm{Cov}(X, \widehat{X}_{det}|R_X=1)\Pr(R_X=1)\}\\
=&\textrm{Var}(X) + \textrm{Var}(X|R_X=1)\Pr(R_X=1)\\
&-R^2\textrm{Var}(X|R_X=0)\frac{\textrm{Var}(Z|R_X=1)}{\textrm{Var}(Z|R_X=0)}\Pr(R_X=1) \\
&-2\{\textrm{Var}(X|R_X=1)\Pr(R_X=1) -\textrm{Cov}(X, \widehat\alpha_0+\widehat\alpha_1Z|R_X=1)\Pr(R_X=1)\}\\
=&\textrm{Var}(X) + \textrm{Var}(X|R_X=1)\Pr(R_X=1)\\
&-R^2\textrm{Var}(X|R_X=0)\frac{\textrm{Var}(Z|R_X=1)}{\textrm{Var}(Z|R_X=0)}\Pr(R_X=1) \\
&-2\textrm{Var}(X|R_X=1)\Pr(R_X=1)\\
&+2R^2\textrm{Var}(X|R_X=0)\underbrace{\frac{\textrm{Cov}(X,Z|R_X=1)}{\textrm{Cov}(X,Z|R_X=0)}}_{\textrm{Cov}(X,Z|R_X) = \textrm{Var}(Z|R_X)}\Pr(R_X=1)\\
=&\textrm{Var}(X)-\textrm{Var}(X|R_X=1)\Pr(R_X=1)\\
&+R^2\textrm{Var}(X|R_X=0)\frac{\textrm{Var}(Z|R_X=1)}{\textrm{Var}(Z|R_X=0)}\Pr(R_X=1)\\
=&\textrm{Var}(X) + \Pr(R_X=1)\left
\{R^2\delta_0\textrm{Var}(X)\frac{\textrm{Var}(Z|R_X=1)}{\textrm{Var}(Z|R_X=0)}-\delta_1\textrm{Var}(X)\right\}\\
=&\textrm{Var}(X)\left[1 + \Pr(R_X=1)\left\{\left(R^2\delta_0\right)\frac{\textrm{Var}(Z|R_X=1)}{\textrm{Var}(Z|R_X=0)}-\delta_1\right\}\right],
\end{aligned}
\]

where we define \(\delta_0\textrm{Var}(X) =\textrm{Var}(X|R_X=0)\), and
\(\delta_1\textrm{Var}(X)=\textrm{Var}(X|R_X=1)\). We can also write
\(\textrm{Cov}(X_{imp,det},Y)\) as a function of \(\textrm{Cov}(X,Y)\)
as follows:

\[
\begin{aligned}
\textrm{Cov}&(X_{imp,det}, Y) = \textrm{Cov}(X-R_X\widehat\varepsilon_X, Y) \\
=&\textrm{Cov}(X,Y) - \textrm{Cov}(R_X\widehat\varepsilon_X, Y)\\
=&\textrm{Cov}(X,Y)- \textrm{Cov}(R_X(X-\widehat{X}_{det}), Y)\\
=&\textrm{Cov}(X,Y)- \textrm{Cov}(R_X X, Y)+\textrm{Cov}(R_X\widehat{X}_{det}, Y)\\
=&\textrm{Cov}(X,Y)\\
&- \Pr(R_X=1)\Pr(R_X=0)\textrm{E}(X|R_X=1)\{\textrm{E}(Y|R_X=1)- \textrm{E}(Y|R_X=0)\}\\
&- \textrm{Cov}(X, Y|R_X=1)\Pr(R_X=1)\\
&+\Pr(R_X=1)\Pr(R_X=0)\textrm{E}(\widehat{X}_{det}|R_X=1)\{\textrm{E}(Y|R_X=1) - \textrm{E}(Y|R_X=0)\}\\
&+\textrm{Cov}(\widehat{X}_{det}, Y|R_X=1)\Pr(R_X=1)\\
=&\textrm{Cov}(X,Y)+ \textrm{Cov}(X, Y|R_X=1)\Pr(R_X=1)\\
&+\textrm{Cov}(\widehat{X}_{det}, Y|R_X=1)\Pr(R_X=1)\\
=&\textrm{Cov}(X,Y)+\Pr(R_X=1)\{\textrm{Cov}(\widehat{X}_{det}, Y|R_X=1)-\textrm{Cov}(X, Y|R_X=1)\}\\
=&\textrm{Cov}(X,Y)+\Pr(R_X=1)\{\textrm{Cov}(\widehat\alpha_0+\widehat\alpha_1Z, Y|R_X=1)-\textrm{Cov}(X, Y|R_X=1)\}\\
=&\textrm{Cov}(X,Y)+\Pr(R_X=1)\{\widehat\alpha_1\textrm{Cov}(Z, Y|R_X=1)-\textrm{Cov}(X, Y|R_X=1)\}\\
=&\textrm{Cov}(X,Y)\\
&+\Pr(R_X=1)\left\{\frac{R^2\textrm{Var}(X|R_X=0)}{\textrm{Cov}(X,Z|R_X=0)}\textrm{Cov}(Z, Y|R_X=1)-\textrm{Cov}(X, Y|R_X=1)\right\}\\
=&\textrm{Cov}(X,Y)\\
&+\Pr(R_X=1)\biggl\{\frac{\beta_1R^2\delta_0\textrm{Var}(X)^2}{\textrm{Cov}(X,Z|R_X=0)\textrm{Cov}(X,Y)}\textrm{Cov}(Z, Y|R_X=1)\\
&\hspace{2.75cm}-\textrm{Cov}(X,Y|R_X=1)\frac{\textrm{Var}(X|R_X=1)}{\textrm{Var}(X|R_X=1)}\biggr\}\\
=&\textrm{Cov}(X,Y)\\
&+\Pr(R_X=1)\left[\beta_1\left\{\frac{R^2\delta_0\textrm{Var}(X)^2}{\textrm{Cov}(X,Z|R_X=0)\textrm{Cov}(X,Y)}\textrm{Cov}(Z, Y|R_X=1)-\delta_1\textrm{Var}(X)\right\}\right]\\
=&\textrm{Cov}(X,Y)\\
&+\Pr(R_X=1)\left[\textrm{Cov}(X,Y)\left\{\frac{R^2\delta_0\textrm{Var}(X)}{\textrm{Cov}(X,Z|R_X=0)\textrm{Cov}(X,Y)}\textrm{Cov}(Z, Y|R_X=1)-\delta_1\right\}\right]\\
=&\textrm{Cov}(X,Y)\\
&+\Pr(R_X=1)\left[\textrm{Cov}(X,Y)\left\{\frac{R^2\delta_0}{\textrm{Var}(Z|R_X=0)\alpha_1 \beta_1}\frac{\textrm{Cov}(Z, Y|R_X=1)}{\textrm{Var}(Z|R_X=1)}\textrm{Var}(Z|R_X=1)-\delta_1\right\}\right]\\
=&\textrm{Cov}(X,Y)\\
&+\Pr(R_X=1)\left[\textrm{Cov}(X,Y)\left\{\frac{R^2\delta_0}{\textrm{Var}(Z|R_X=0)}\textrm{Var}(Z|R_X=1)-\delta_1\right\}\right]\\
=&\textrm{Cov}(X,Y)\left[1+\Pr(R_X=1)\left\{\frac{R^2\delta_0}{\textrm{Var}(Z|R_X=0)}\textrm{Var}(Z|R_X=1)-\delta_1\right\}\right].\\
\end{aligned}
\]

Note that:

\[
\begin{aligned}
\frac{\textrm{Cov}(Z, Y)}{\textrm{Var}(Z)}& = \frac{\textrm{Cov}(X,Z)\textrm{Cov}(X,Y)}{\textrm{Var}(Z)\textrm{Var}(X)}\\
\frac{\textrm{Cov}(Z, \beta_0+\beta_1X+\varepsilon_Y)}{\textrm{Var}(Z)}&=\alpha_1\beta_1\\
\frac{\beta_1\textrm{Cov}(Z,X) + \textrm{Cov}(Z,\varepsilon_Y)}{\textrm{Var}(Z)}&=\alpha_1\beta_1\\
\frac{\beta_1\textrm{Cov}(Z,X)}{\textrm{Var}(Z)}&=\alpha_1\beta_1\\
\beta_1\alpha_1&=\alpha_1\beta_1.
\end{aligned}
\]

Therefore, \(\textrm{Var}(X)\omega = \textrm{Var}(X_{imp,det})\) and
\(\textrm{Cov}(X,Y)\omega = \textrm{Cov}(X_{imp,det}, Y)\) where
\(\omega = \left[1 + \Pr(R_X=1)\left\{R^2\delta_0\left(\frac{\textrm{Var}(Z|R_X=1)}{\textrm{Var}(Z|R_X=0)}\right)-\delta_1\right\}\right]\)
where \(R^2\) is the \(R^2\) from the imputation model.

\hypertarget{covariance-equivalence-from-deterministic-or-stochastic-imputation}{%
\subsection*{Covariance equivalence from deterministic or stochastic
imputation}\label{covariance-equivalence-from-deterministic-or-stochastic-imputation}}
\addcontentsline{toc}{subsection}{Covariance equivalence from
deterministic or stochastic imputation}

The derivation below shows that the covariance between the imputed \(X\)
and \(Y\) after imputing using a stochastic process is the same as that
using a deterministic process when using the same imputation model.
Here, the deterministic imputation imputes as follows:

\[
X_{imp,det} =\begin{cases}
X_{obs} &\textrm{ if }x\textrm{ is not missing}\\
\widehat\alpha_0 + \widehat\alpha_1z &\textrm{ otherwise}.
\end{cases}
\]

The stochastic model imputes as follows:

\[
X_{imp, stoc} =\begin{cases}
X_{obs} &\textrm{ if }x\textrm{ is not missing}\\
\dot\alpha_0 + \dot\alpha_1z +\dot\varepsilon&\textrm{ otherwise}
\end{cases}
\]

where \(\dot{\varepsilon} = N(0, \dot\sigma^2)\) and \(\dot\sigma^2\) is
drawn from a scaled inverse-chi squared distribution with a scale
parameter equal to the mean residual sum of squares from the
Equation~\ref{eq-1} with \(n_{obs}-2\) degrees of freedom. Parameters
\(\dot{\alpha}_0\) and \(\dot{\alpha}_1\) are drawn from a bivariate
normal distribution with a mean of
(\(\widehat\alpha_0, \widehat\alpha_1\)) and a variance-covariance
matrix taken from the observed variance-covariance matrix from the
imputation model multiplied by \(\dot\sigma^2\). We could equivalently
represent the stochastic parameters as
\(\dot\alpha_0=\widehat\alpha_0+u_0\) and
\(\dot\alpha_1=\widehat\alpha_1+u_1\), where \(u_0\) and \(u_1\)
represent the uncertainty as described.

Using this framing, the following demonstrates that the covariances
between \(Y\) and either imputed value (\(X_{imp,det}\) and
\(X_{imp,stoc}\)) are equivalent.

\[
\begin{aligned}
\textrm{Cov}&(X_{imp,stoc}, Y) = \textrm{E}(X_{imp,stoc}Y)-\textrm{E}(X_{imp,stoc})\textrm{E}(Y)\\
=&  \textrm{E}(X_{imp,stoc}Y)-\textrm{E}(X_{imp,det})\textrm{E}(Y) \hspace{2cm} \textrm{because }\textrm{E}(X_{imp,stoc})=\textrm{E}(X_{imp,det})\\
=&\textrm{E}\{XY+R_X(X-\widehat{X}_{stoc})Y\}-\textrm{E}(X_{imp,det})\textrm{E}(Y)\\
=&\textrm{E}(XY)+\textrm{E}\{R_X(XY)-R_X(\widehat{X}_{stoc}Y)\}-\textrm{E}(X_{imp,det})\textrm{E}(Y)\\
=&\textrm{E}(XY)+\textrm{E}\{R_X(XY)\}-\textrm{E}\{R_X(\widehat{X}_{stoc}Y)\}-\textrm{E}(X_{imp,det})\textrm{E}(Y)\\
=&\textrm{E}(XY)+\textrm{E}\{R_X(XY)\}-\textrm{E}\{R_X(\dot\alpha_0Y) + R_X(\dot\alpha_1ZY)+R_X(\dot\varepsilon Y)\}-\textrm{E}(X_{imp,det})\textrm{E}(Y)\\
=&\textrm{E}(XY)+\textrm{E}\{R_X(XY)\}-\textrm{E}\{R_X(\widehat\alpha_0+u_0)Y)\} - \textrm{E}\{(R_X(\widehat\alpha_1+u_1)ZY\}-\textrm{E}(R_X(\dot\varepsilon Y))\\
&-\textrm{E}(X_{imp,det})\textrm{E}(Y)\\
=&\textrm{E}(XY)+\textrm{E}\{R_X(XY)\}-\widehat\alpha_0\textrm{E}(R_XY)-\textrm{E}(u_0)\textrm{E}(R_XY) - \widehat\alpha_1\textrm{E}(R_XZY)\\
&-\textrm{E}(u_1)\textrm{E}(R_XZY)-\textrm{E}(R_X(\dot\varepsilon Y))-\textrm{E}(X_{imp,det})\textrm{E}(Y)\\
=&\textrm{E}(XY)+\textrm{E}\{R_X(XY)\}-\widehat\alpha_0\textrm{E}(R_XY)- \widehat\alpha_1\textrm{E}(R_XZY)-\textrm{E}(R_X(\dot\varepsilon Y))-\textrm{E}(X_{imp,det})\textrm{E}(Y)\\
=&\textrm{E}(XY)+\textrm{E}\{R_X(XY)\}-\widehat\alpha_0\textrm{E}(R_XY)- \widehat\alpha_1\textrm{E}(R_XZY)-\textrm{E}(X_{imp,det})\textrm{E}(Y)\\
=&\textrm{E}\{XY + R_X(X-\widehat{X}_{det})Y\}-\textrm{E}(X_{imp,det})\textrm{E}(Y)\\
=& \textrm{Cov}(X_{imp,det}, Y)
\end{aligned}
\]

\hypertarget{appendix-b---simulation}{%
\section*{Appendix B - Simulation}\label{appendix-b---simulation}}
\addcontentsline{toc}{section}{Appendix B - Simulation}

Simulations were conducted using R version 4.2 (31,32). The code used to
run the simulations and create the figures can be found below. To
demonstrate the results that we derived for a wider range of examples,
we simulated nine large data sets, each with \(n = 1{,}000{,}000\)
observations. We simulated \(X\) and \(Z\) as described in
Equation~\ref{eq-xz} and the relationship between \(X\) and \(Y\) as
described in Equation~\ref{eq-y}. We varied the proportion of missing
values as well as the magnitude of differential missingness under nine
scenarios based on

\[
\Pr(R_X=1|Z) = 0.25(1-Z) + p(Z),
\]

where \(p\) varied from 0.05 to 0.85 by 0.1. In other words, the
probability that the observed \(x\) was missing was always 0.25 when
\(Z=0\) (as throughout Section~\ref{sec-methods}), and the probability
that \(X\) was missing when \(Z=1\) ranged from 0.05 to 0.85. Notably,
when \(p=0.25\), the observed \(x\) was missing completely at random
since \(\Pr(R_X=1|Z)=\Pr(R_X=1)=0.25\).

Using the simulated data, we fit four imputation models: (i) a
deterministic model without the outcome, (ii) a deterministic model with
the outcome, (iii) a stochastic model without the outcome, and (iv) a
stochastic model with the outcome. Figure~\ref{fig-imp} displays the
results, which support our theoretical derivations previously described.
Figure~\ref{fig-imp-1} shows the deterministic imputation model without
\(Y\); in all scenarios, the variance and covariance are both
underestimated. Despite this underestimation, however, the estimated
coefficient, \(\widehat\beta_{1, Y\sim X_{imp, det}}\) for the
relationship between \(X_{imp, det}\) and \(Y\) is unbiased in all
cases. Figure~\ref{fig-imp-2} shows that conditioning on \(Y\) in the
deterministic outcome model results in the correct estimation of the
covariance but an underestimation of the variance of \(X\), resulting in
a biased estimate, \(\widehat\beta_{1, Y\sim X_{imp,det|y}}\), that
overestimates the true relationship. Figure~\ref{fig-imp-3} demonstrates
that the stochastic imputation model that does not condition on \(Y\)
correctly estimates the variance of \(X\) but underestimates the
covariance between \(X\) and \(Y\), resulting in biased
\(\widehat\beta_{1, Y\sim X_{imp, stoc}}\) estimates that underestimate
the true relationship, \(\beta_1\). For estimates that are biased, the
severity of the bias increased as the magnitude of missingness increased
(i.e., the value of \(p\) is farther from 0). Notably, even when the
data are missing completely at random (i.e., when \(p = 0.25\)) the
estimate for \(\widehat\beta_{1, Y\sim X_{imp, stoc}}\) is quite biased.
Figure~\ref{fig-imp-4} demonstrates that when the outcome is included in
the stochastic imputation model both the covariance and variance are
recovered, resulting in and unbiased estimate,
\(\widehat\beta_{1, Y\sim X_{imp, stoc|y}}\), of the true relationship
between \(X\) and \(Y\).

\spacingset{1}

\begin{figure}

\begin{minipage}[t]{0.48\linewidth}

{\centering 

\raisebox{-\height}{

\includegraphics{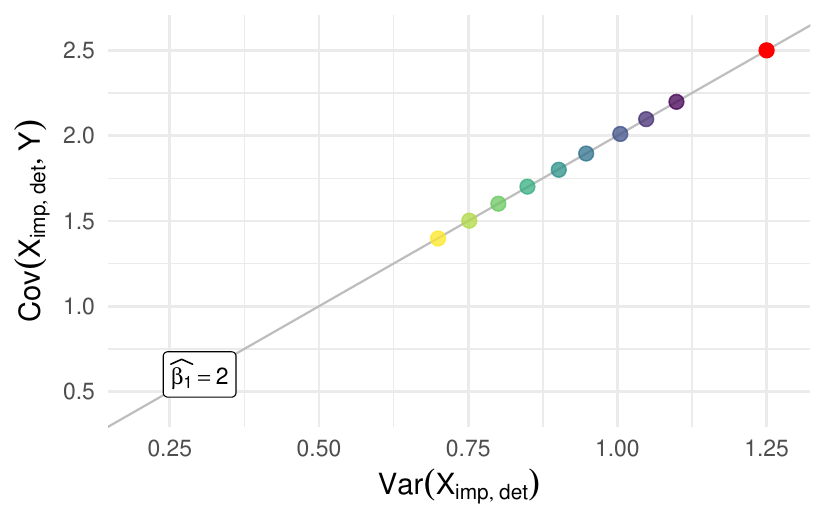}

}

}

\subcaption{\label{fig-imp-1}Deterministic imputation model without
\(Y\). Even though the variance and covariance are both always
underestimated, the ratio of the two (i.e., the parameter estimate
\(\widehat{\beta}_1\)) is always correct.}
\end{minipage}%
\begin{minipage}[t]{0.05\linewidth}

{\centering 

~

}

\end{minipage}%
\begin{minipage}[t]{0.48\linewidth}

{\centering 

\raisebox{-\height}{

\includegraphics{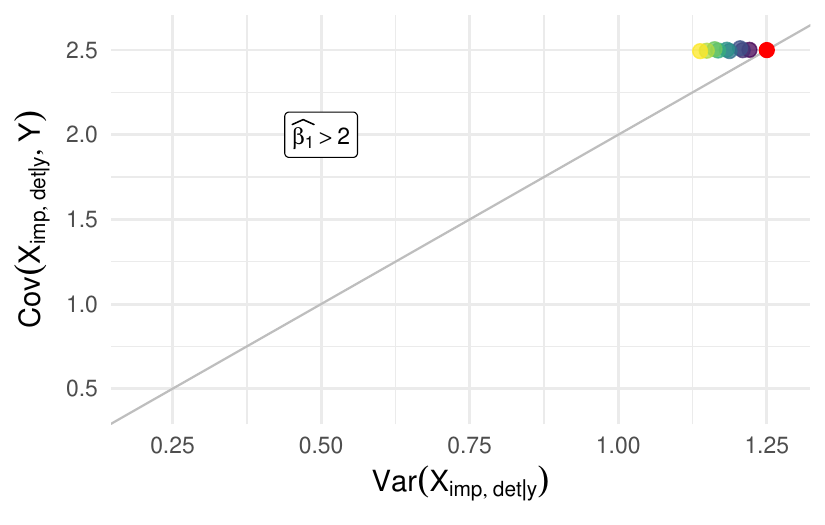}

}

}

\subcaption{\label{fig-imp-2}Deterministic imputation model with \(Y\).
The points always over-estimated the effect between \(X_{imp}\) and
\(Y\) (that is they fall above the grey line). Notice the covariance is
always correctly estimated, but the variance is underestimated (with
maximum bias when the probability of missingness given \(Z=1\) is the
highest), yielding a biased result.}
\end{minipage}%
\newline
\begin{minipage}[t]{0.48\linewidth}

{\centering 

\raisebox{-\height}{

\includegraphics{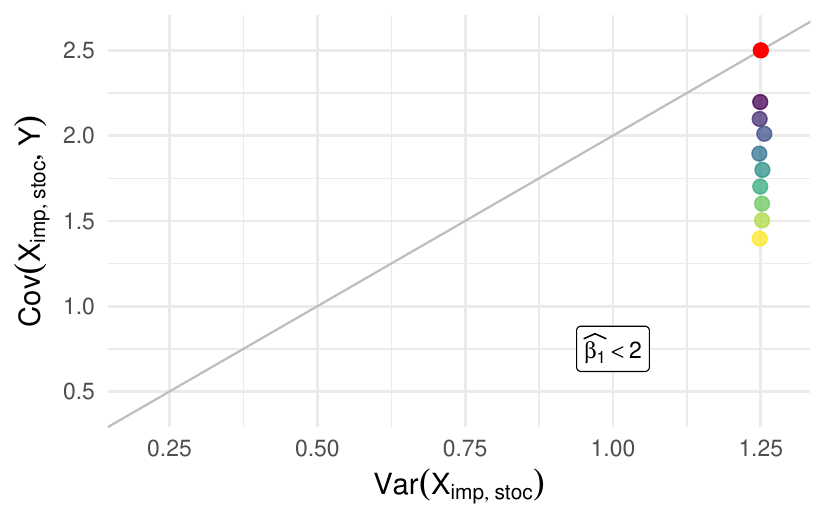}

}

}

\subcaption{\label{fig-imp-3}Stochastic imputation model without \(Y\).
The variance is correctly estimated, but the covariance is still
underestimated (with the degree of underestimation increasing as the
probability that \(X\) is missing given \(Z=1\) increases), yielding a
biased result.}
\end{minipage}%
\begin{minipage}[t]{0.05\linewidth}

{\centering 

~

}

\end{minipage}%
\begin{minipage}[t]{0.48\linewidth}

{\centering 

\raisebox{-\height}{

\includegraphics{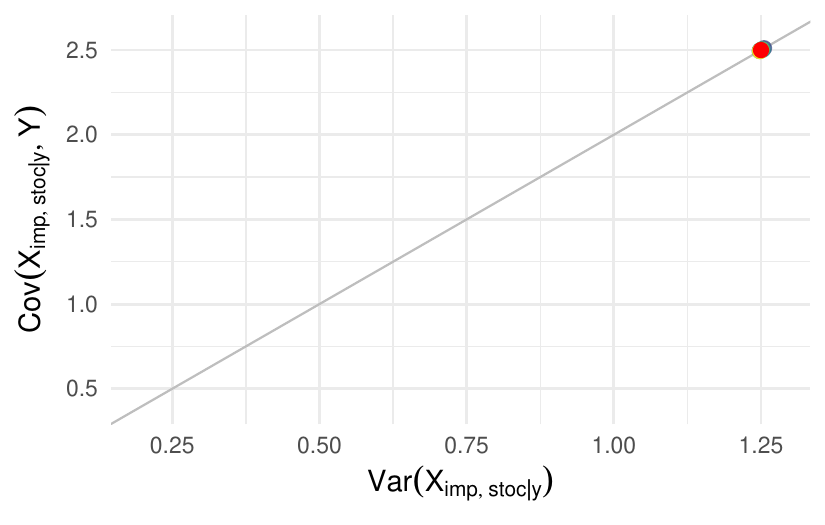}

}

}

\subcaption{\label{fig-imp-4}Stochastic imputation model with \(Y\). The
correct ratio is recovered when the stochastic imputation model includes
\(Y\). Both the variance and the covariance are equal to the true values
if \(X\) were fully observed.}
\end{minipage}%
\newline
\begin{minipage}[t]{\linewidth}

{\centering 

\raisebox{-\height}{

\includegraphics[width=8in,height=\textheight]{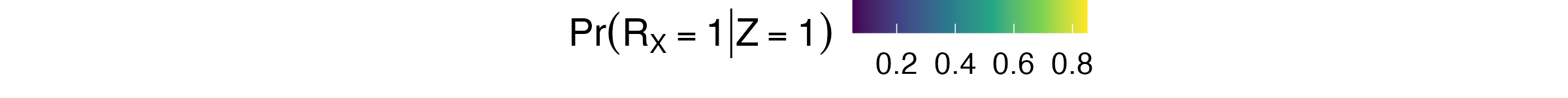}

}

}

\end{minipage}%

\caption{\label{fig-imp}The probability that X is missing given \(Z=0\)
is fixed at 0.25. The colors of the points correspond to the probability
that X is missing given \(Z=1\). The x-axis shows the variance of
\(X_{imp}\) (i.e., the imputed values along with the observed values of
\(X\)). The y-axis shows the covariance between \(X_{imp}\) and \(Y\).
The red dot shows the correct variance and covariance if \(X\) were
fully observed. The grey line shows where the ratio (i.e., the
coefficient estimate) of the two values is equal to the true value of
2.}

\end{figure}

Below is the R code used for the above simulations and to create the
figures in the manuscript.

\begin{Shaded}
\begin{Highlighting}[]
\FunctionTok{library}\NormalTok{(tidyverse)}
\DocumentationTok{\#\# Generate data for simulation}

\FunctionTok{set.seed}\NormalTok{(}\DecValTok{1}\NormalTok{)}
\NormalTok{n }\OtherTok{\textless{}{-}} \DecValTok{1000000}
\NormalTok{sym }\OtherTok{\textless{}{-}} \ControlFlowTok{function}\NormalTok{(x) \{}
\NormalTok{  (x }\SpecialCharTok{+} \FunctionTok{t}\NormalTok{(x)) }\SpecialCharTok{/} \DecValTok{2}
\NormalTok{\}}
\NormalTok{gen\_data }\OtherTok{\textless{}{-}} \ControlFlowTok{function}\NormalTok{(p\_miss\_1) \{}
\NormalTok{  data }\OtherTok{\textless{}{-}} \FunctionTok{tibble}\NormalTok{(}
    \AttributeTok{z =} \FunctionTok{rbinom}\NormalTok{(n, }\DecValTok{1}\NormalTok{, }\AttributeTok{p =} \FloatTok{0.5}\NormalTok{),}
    \AttributeTok{x =}\NormalTok{ z }\SpecialCharTok{+} \FunctionTok{rnorm}\NormalTok{(n),}
    \AttributeTok{y =} \DecValTok{2} \SpecialCharTok{*}\NormalTok{ x }\SpecialCharTok{+} \FunctionTok{rnorm}\NormalTok{(n),}
    \AttributeTok{x\_miss =} \FunctionTok{ifelse}\NormalTok{(z, }\FunctionTok{rbinom}\NormalTok{(n, }\DecValTok{1}\NormalTok{, p\_miss\_1), }\FunctionTok{rbinom}\NormalTok{(n, }\DecValTok{1}\NormalTok{, }\FloatTok{0.25}\NormalTok{)),}
    \AttributeTok{x\_obs =} \FunctionTok{ifelse}\NormalTok{(x\_miss, }\ConstantTok{NA}\NormalTok{, x),}
    \AttributeTok{p\_miss\_1 =}\NormalTok{ p\_miss\_1}
\NormalTok{  )}
\NormalTok{  imp\_fit }\OtherTok{\textless{}{-}} \FunctionTok{lm}\NormalTok{(x\_obs }\SpecialCharTok{\textasciitilde{}}\NormalTok{ z, }\AttributeTok{data =}\NormalTok{ data)}
\NormalTok{  coefs }\OtherTok{\textless{}{-}} \FunctionTok{coef}\NormalTok{(imp\_fit)}
\NormalTok{  sigma }\OtherTok{\textless{}{-}} \FunctionTok{sqrt}\NormalTok{(}\FunctionTok{sum}\NormalTok{(imp\_fit}\SpecialCharTok{$}\NormalTok{residuals}\SpecialCharTok{\^{}}\DecValTok{2}\NormalTok{) }\SpecialCharTok{/} \FunctionTok{rchisq}\NormalTok{(}\DecValTok{1}\NormalTok{, imp\_fit}\SpecialCharTok{$}\NormalTok{df))}
\NormalTok{  beta }\OtherTok{\textless{}{-}}\NormalTok{  coefs }\SpecialCharTok{+} \FunctionTok{t}\NormalTok{(}\FunctionTok{chol}\NormalTok{(}\FunctionTok{sym}\NormalTok{(}\FunctionTok{vcov}\NormalTok{(imp\_fit)))) }\SpecialCharTok{\%*\%} \FunctionTok{rnorm}\NormalTok{(}\DecValTok{2}\NormalTok{, }\DecValTok{0}\NormalTok{, sigma) }
\NormalTok{  x\_p }\OtherTok{\textless{}{-}}\NormalTok{ beta[}\DecValTok{1}\NormalTok{] }\SpecialCharTok{+}\NormalTok{ beta[}\DecValTok{2}\NormalTok{] }\SpecialCharTok{*}\NormalTok{ data}\SpecialCharTok{$}\NormalTok{z  }\SpecialCharTok{+} \FunctionTok{rnorm}\NormalTok{(n, }\DecValTok{0}\NormalTok{, sigma)}
  
\NormalTok{  imp\_fit }\OtherTok{\textless{}{-}} \FunctionTok{lm}\NormalTok{(x\_obs }\SpecialCharTok{\textasciitilde{}}\NormalTok{ z }\SpecialCharTok{+}\NormalTok{ y, }\AttributeTok{data =}\NormalTok{ data)}
\NormalTok{  coefs }\OtherTok{\textless{}{-}} \FunctionTok{coef}\NormalTok{(imp\_fit)}
\NormalTok{  sigma }\OtherTok{\textless{}{-}} \FunctionTok{sqrt}\NormalTok{(}\FunctionTok{sum}\NormalTok{(imp\_fit}\SpecialCharTok{$}\NormalTok{residuals}\SpecialCharTok{\^{}}\DecValTok{2}\NormalTok{) }\SpecialCharTok{/} \FunctionTok{rchisq}\NormalTok{(}\DecValTok{1}\NormalTok{, imp\_fit}\SpecialCharTok{$}\NormalTok{df))}
\NormalTok{  beta }\OtherTok{\textless{}{-}}\NormalTok{  coefs }\SpecialCharTok{+} \FunctionTok{t}\NormalTok{(}\FunctionTok{chol}\NormalTok{(}\FunctionTok{sym}\NormalTok{(}\FunctionTok{vcov}\NormalTok{(imp\_fit)))) }\SpecialCharTok{\%*\%} \FunctionTok{rnorm}\NormalTok{(}\DecValTok{3}\NormalTok{, }\DecValTok{0}\NormalTok{, sigma) }
\NormalTok{  x\_p\_y }\OtherTok{\textless{}{-}}\NormalTok{ beta[}\DecValTok{1}\NormalTok{] }\SpecialCharTok{+}\NormalTok{ beta[}\DecValTok{2}\NormalTok{] }\SpecialCharTok{*}\NormalTok{ data}\SpecialCharTok{$}\NormalTok{z }\SpecialCharTok{+}\NormalTok{ beta[}\DecValTok{3}\NormalTok{] }\SpecialCharTok{*}\NormalTok{ data}\SpecialCharTok{$}\NormalTok{y }\SpecialCharTok{+} \FunctionTok{rnorm}\NormalTok{(n, }\DecValTok{0}\NormalTok{, sigma)}
\NormalTok{  data }\SpecialCharTok{|\textgreater{}}
    \FunctionTok{mutate}\NormalTok{(}\AttributeTok{x\_imp =} \FunctionTok{ifelse}\NormalTok{(x\_miss, }
                          \FunctionTok{predict}\NormalTok{(}\FunctionTok{lm}\NormalTok{(x\_obs }\SpecialCharTok{\textasciitilde{}}\NormalTok{ z, }\AttributeTok{data =}\NormalTok{ data), data),}
\NormalTok{                          x\_obs),}
           \AttributeTok{x\_impy =} \FunctionTok{ifelse}\NormalTok{(x\_miss, }
                           \FunctionTok{predict}\NormalTok{(}\FunctionTok{lm}\NormalTok{(x\_obs }\SpecialCharTok{\textasciitilde{}}\NormalTok{ z }\SpecialCharTok{+}\NormalTok{ y, }\AttributeTok{data =}\NormalTok{ data), data),}
\NormalTok{                           x\_obs),}
           \AttributeTok{x\_impmi =} \FunctionTok{ifelse}\NormalTok{(x\_miss, x\_p, x\_obs),}
           \AttributeTok{x\_impmiy =} \FunctionTok{ifelse}\NormalTok{(x\_miss, x\_p\_y, x\_obs)}
\NormalTok{    )}
\NormalTok{\}}

\NormalTok{d }\OtherTok{\textless{}{-}} \FunctionTok{map\_df}\NormalTok{(}\FunctionTok{seq}\NormalTok{(}\FloatTok{0.05}\NormalTok{, }\FloatTok{0.9}\NormalTok{, }\AttributeTok{by =} \FloatTok{0.1}\NormalTok{), gen\_data)}
\DocumentationTok{\#\# Create Figure 2}
\NormalTok{d }\SpecialCharTok{|\textgreater{}}
  \FunctionTok{group\_by}\NormalTok{(p\_miss\_1) }\SpecialCharTok{|\textgreater{}}
  \FunctionTok{summarise}\NormalTok{(}\AttributeTok{cov\_imp =} \FunctionTok{cov}\NormalTok{(x\_imp, y),}
            \AttributeTok{var\_imp =} \FunctionTok{var}\NormalTok{(x\_imp),}
            \AttributeTok{beta\_imp =}\NormalTok{ cov\_imp }\SpecialCharTok{/}\NormalTok{ var\_imp,}
            \AttributeTok{cov\_impy =} \FunctionTok{cov}\NormalTok{(x\_impy, y),}
            \AttributeTok{var\_impy =} \FunctionTok{var}\NormalTok{(x\_impy),}
            \AttributeTok{beta\_impy =}\NormalTok{ cov\_impy }\SpecialCharTok{/}\NormalTok{ var\_impy,}
            \AttributeTok{cov\_impmi =} \FunctionTok{cov}\NormalTok{(x\_impmi, y),}
            \AttributeTok{var\_impmi =} \FunctionTok{var}\NormalTok{(x\_impmi),}
            \AttributeTok{beta\_impmi =}\NormalTok{ cov\_impmi }\SpecialCharTok{/}\NormalTok{ var\_impmi,}
            \AttributeTok{cov\_impmiy =} \FunctionTok{cov}\NormalTok{(x\_impmiy, y),}
            \AttributeTok{var\_impmiy =} \FunctionTok{var}\NormalTok{(x\_impmiy),}
            \AttributeTok{beta\_impmiy =}\NormalTok{ cov\_impmiy }\SpecialCharTok{/}\NormalTok{ var\_impmiy) }\OtherTok{{-}\textgreater{}}\NormalTok{ sum\_d}


\FunctionTok{ggplot}\NormalTok{(sum\_d, }\FunctionTok{aes}\NormalTok{(}\AttributeTok{x =}\NormalTok{ var\_imp,  }\AttributeTok{y =}\NormalTok{ cov\_imp, }\AttributeTok{color =}\NormalTok{ p\_miss\_1)) }\SpecialCharTok{+}
  \FunctionTok{geom\_abline}\NormalTok{(}\AttributeTok{slope =} \DecValTok{2}\NormalTok{, }\AttributeTok{color =} \StringTok{"grey"}\NormalTok{)}\SpecialCharTok{+}
  \FunctionTok{geom\_point}\NormalTok{(}\AttributeTok{alpha =} \FloatTok{0.75}\NormalTok{) }\SpecialCharTok{+}
  \FunctionTok{geom\_point}\NormalTok{(}\AttributeTok{data =} \FunctionTok{data.frame}\NormalTok{(}
    \AttributeTok{var =} \FloatTok{1.25}\NormalTok{, }\AttributeTok{cov =} \FloatTok{2.5}
\NormalTok{  ), }\FunctionTok{aes}\NormalTok{(}\AttributeTok{x =}\NormalTok{ var, }\AttributeTok{y =}\NormalTok{ cov), }\AttributeTok{color =} \StringTok{"red"}\NormalTok{) }\SpecialCharTok{+}
  \FunctionTok{labs}\NormalTok{(}\AttributeTok{x =}\NormalTok{ latex2exp}\SpecialCharTok{::}\FunctionTok{TeX}\NormalTok{(}\StringTok{"$}\SpecialCharTok{\textbackslash{}\textbackslash{}}\StringTok{textrm\{Var\}(X\_\{imp,det\})$"}\NormalTok{),}
       \AttributeTok{y =}\NormalTok{ latex2exp}\SpecialCharTok{::}\FunctionTok{TeX}\NormalTok{(}\StringTok{"$}\SpecialCharTok{\textbackslash{}\textbackslash{}}\StringTok{textrm\{Cov\}(X\_\{imp,det\}, Y)$"}\NormalTok{)) }\SpecialCharTok{+}
  \FunctionTok{xlim}\NormalTok{(}\FunctionTok{c}\NormalTok{(}\FloatTok{0.2}\NormalTok{, }\FloatTok{1.27}\NormalTok{)) }\SpecialCharTok{+}
  \FunctionTok{ylim}\NormalTok{(}\FunctionTok{c}\NormalTok{(}\FloatTok{0.4}\NormalTok{, }\FloatTok{2.6}\NormalTok{)) }\SpecialCharTok{+}
  \FunctionTok{annotate}\NormalTok{(}\StringTok{"label"}\NormalTok{, }\AttributeTok{x =} \FloatTok{0.3}\NormalTok{, }\AttributeTok{y =} \FloatTok{0.6}\NormalTok{, }
           \AttributeTok{label =}\NormalTok{ latex2exp}\SpecialCharTok{::}\FunctionTok{TeX}\NormalTok{(}\StringTok{"$}\SpecialCharTok{\textbackslash{}\textbackslash{}}\StringTok{widehat\{}\SpecialCharTok{\textbackslash{}\textbackslash{}}\StringTok{beta\_1\}=2$"}\NormalTok{)) }\SpecialCharTok{+}
  \FunctionTok{theme}\NormalTok{(}\AttributeTok{legend.position =} \StringTok{"none"}\NormalTok{)}

\FunctionTok{ggplot}\NormalTok{(sum\_d, }\FunctionTok{aes}\NormalTok{(}\AttributeTok{x =}\NormalTok{ var\_impy,  }\AttributeTok{y =}\NormalTok{ cov\_impy, }\AttributeTok{color =}\NormalTok{ p\_miss\_1)) }\SpecialCharTok{+}
  \FunctionTok{geom\_abline}\NormalTok{(}\AttributeTok{slope =} \DecValTok{2}\NormalTok{, }\AttributeTok{color =} \StringTok{"grey"}\NormalTok{)}\SpecialCharTok{+}
  \FunctionTok{geom\_point}\NormalTok{(}\AttributeTok{alpha =} \FloatTok{0.75}\NormalTok{) }\SpecialCharTok{+}
  \FunctionTok{geom\_point}\NormalTok{(}\AttributeTok{data =} \FunctionTok{data.frame}\NormalTok{(}
    \AttributeTok{var =} \FloatTok{1.25}\NormalTok{, }\AttributeTok{cov =} \FloatTok{2.5}
\NormalTok{  ), }\FunctionTok{aes}\NormalTok{(}\AttributeTok{x =}\NormalTok{ var, }\AttributeTok{y =}\NormalTok{ cov), }\AttributeTok{color =} \StringTok{"red"}\NormalTok{) }\SpecialCharTok{+}
  \FunctionTok{labs}\NormalTok{(}\AttributeTok{x =}\NormalTok{ latex2exp}\SpecialCharTok{::}\FunctionTok{TeX}\NormalTok{(}\StringTok{"$}\SpecialCharTok{\textbackslash{}\textbackslash{}}\StringTok{textrm\{Var\}(X\_\{imp,det|y\})$"}\NormalTok{),}
       \AttributeTok{y =}\NormalTok{ latex2exp}\SpecialCharTok{::}\FunctionTok{TeX}\NormalTok{(}\StringTok{"$}\SpecialCharTok{\textbackslash{}\textbackslash{}}\StringTok{textrm\{Cov\}(X\_\{imp,det|y\}, Y)$"}\NormalTok{)) }\SpecialCharTok{+}
  \FunctionTok{xlim}\NormalTok{(}\FunctionTok{c}\NormalTok{(}\FloatTok{0.2}\NormalTok{, }\FloatTok{1.27}\NormalTok{)) }\SpecialCharTok{+}
  \FunctionTok{ylim}\NormalTok{(}\FunctionTok{c}\NormalTok{(}\FloatTok{0.4}\NormalTok{, }\FloatTok{2.6}\NormalTok{)) }\SpecialCharTok{+}
  \FunctionTok{annotate}\NormalTok{(}\StringTok{"label"}\NormalTok{, }\AttributeTok{x =} \FloatTok{0.5}\NormalTok{, }\AttributeTok{y =} \DecValTok{2}\NormalTok{, }
           \AttributeTok{label =}\NormalTok{ latex2exp}\SpecialCharTok{::}\FunctionTok{TeX}\NormalTok{(}\StringTok{"$}\SpecialCharTok{\textbackslash{}\textbackslash{}}\StringTok{widehat\{}\SpecialCharTok{\textbackslash{}\textbackslash{}}\StringTok{beta\_1\}\textgreater{}2$"}\NormalTok{)) }\SpecialCharTok{+}
  \FunctionTok{theme}\NormalTok{(}\AttributeTok{legend.position =} \StringTok{"none"}\NormalTok{)}


\FunctionTok{ggplot}\NormalTok{(sum\_d, }\FunctionTok{aes}\NormalTok{(}\AttributeTok{x =}\NormalTok{ var\_impmi,  }\AttributeTok{y =}\NormalTok{ cov\_impmi, }\AttributeTok{color =}\NormalTok{ p\_miss\_1)) }\SpecialCharTok{+}
  \FunctionTok{geom\_abline}\NormalTok{(}\AttributeTok{slope =} \DecValTok{2}\NormalTok{, }\AttributeTok{color =} \StringTok{"grey"}\NormalTok{)}\SpecialCharTok{+}
  \FunctionTok{geom\_point}\NormalTok{(}\AttributeTok{alpha =} \FloatTok{0.75}\NormalTok{) }\SpecialCharTok{+}
  \FunctionTok{geom\_point}\NormalTok{(}\AttributeTok{data =} \FunctionTok{data.frame}\NormalTok{(}
    \AttributeTok{var =} \FloatTok{1.25}\NormalTok{, }\AttributeTok{cov =} \FloatTok{2.5}
\NormalTok{  ), }\FunctionTok{aes}\NormalTok{(}\AttributeTok{x =}\NormalTok{ var, }\AttributeTok{y =}\NormalTok{ cov), }\AttributeTok{color =} \StringTok{"red"}\NormalTok{) }\SpecialCharTok{+}
  \FunctionTok{labs}\NormalTok{(}\AttributeTok{x =}\NormalTok{ latex2exp}\SpecialCharTok{::}\FunctionTok{TeX}\NormalTok{(}\StringTok{"$}\SpecialCharTok{\textbackslash{}\textbackslash{}}\StringTok{textrm\{Var\}(X\_\{imp, stoc\})$"}\NormalTok{),}
       \AttributeTok{y =}\NormalTok{ latex2exp}\SpecialCharTok{::}\FunctionTok{TeX}\NormalTok{(}\StringTok{"$}\SpecialCharTok{\textbackslash{}\textbackslash{}}\StringTok{textrm\{Cov\}(X\_\{imp, stoc\}, Y)$"}\NormalTok{)) }\SpecialCharTok{+}
  \FunctionTok{xlim}\NormalTok{(}\FunctionTok{c}\NormalTok{(}\FloatTok{0.2}\NormalTok{, }\FloatTok{1.28}\NormalTok{)) }\SpecialCharTok{+}
  \FunctionTok{ylim}\NormalTok{(}\FunctionTok{c}\NormalTok{(}\FloatTok{0.4}\NormalTok{, }\FloatTok{2.6}\NormalTok{)) }\SpecialCharTok{+}
  \FunctionTok{annotate}\NormalTok{(}\StringTok{"label"}\NormalTok{, }\AttributeTok{x =} \DecValTok{1}\NormalTok{, }\AttributeTok{y =} \FloatTok{0.75}\NormalTok{, }
           \AttributeTok{label =}\NormalTok{ latex2exp}\SpecialCharTok{::}\FunctionTok{TeX}\NormalTok{(}\StringTok{"$}\SpecialCharTok{\textbackslash{}\textbackslash{}}\StringTok{widehat\{}\SpecialCharTok{\textbackslash{}\textbackslash{}}\StringTok{beta\_1\}\textless{}2$"}\NormalTok{)) }\SpecialCharTok{+}
  \FunctionTok{theme}\NormalTok{(}\AttributeTok{legend.position =} \StringTok{"none"}\NormalTok{)}

\FunctionTok{ggplot}\NormalTok{(sum\_d, }\FunctionTok{aes}\NormalTok{(}\AttributeTok{x =}\NormalTok{ var\_impmiy,  }\AttributeTok{y =}\NormalTok{ cov\_impmiy, }\AttributeTok{color =}\NormalTok{ p\_miss\_1)) }\SpecialCharTok{+}
  \FunctionTok{geom\_abline}\NormalTok{(}\AttributeTok{slope =} \DecValTok{2}\NormalTok{, }\AttributeTok{color =} \StringTok{"grey"}\NormalTok{)}\SpecialCharTok{+}
  \FunctionTok{geom\_point}\NormalTok{(}\AttributeTok{alpha =} \FloatTok{0.75}\NormalTok{) }\SpecialCharTok{+}
  \FunctionTok{geom\_point}\NormalTok{(}\AttributeTok{data =} \FunctionTok{data.frame}\NormalTok{(}
    \AttributeTok{var =} \FloatTok{1.25}\NormalTok{, }\AttributeTok{cov =} \FloatTok{2.5}
\NormalTok{  ), }\FunctionTok{aes}\NormalTok{(}\AttributeTok{x =}\NormalTok{ var, }\AttributeTok{y =}\NormalTok{ cov), }\AttributeTok{color =} \StringTok{"red"}\NormalTok{) }\SpecialCharTok{+}
  \FunctionTok{labs}\NormalTok{(}\AttributeTok{x =}\NormalTok{ latex2exp}\SpecialCharTok{::}\FunctionTok{TeX}\NormalTok{(}\StringTok{"$}\SpecialCharTok{\textbackslash{}\textbackslash{}}\StringTok{textrm\{Var\}(X\_\{imp, stoc|y\})$"}\NormalTok{),}
       \AttributeTok{y =}\NormalTok{ latex2exp}\SpecialCharTok{::}\FunctionTok{TeX}\NormalTok{(}\StringTok{"$}\SpecialCharTok{\textbackslash{}\textbackslash{}}\StringTok{textrm\{Cov\}(X\_\{imp, stoc|y\}, Y)$"}\NormalTok{),}
       \AttributeTok{color =}\NormalTok{ latex2exp}\SpecialCharTok{::}\FunctionTok{TeX}\NormalTok{(}\StringTok{"$}\SpecialCharTok{\textbackslash{}\textbackslash{}}\StringTok{Pr(R\_X=1|Z=1)$"}\NormalTok{)) }\SpecialCharTok{+}
  \FunctionTok{xlim}\NormalTok{(}\FunctionTok{c}\NormalTok{(}\FloatTok{0.2}\NormalTok{, }\FloatTok{1.28}\NormalTok{)) }\SpecialCharTok{+}
  \FunctionTok{ylim}\NormalTok{(}\FunctionTok{c}\NormalTok{(}\FloatTok{0.4}\NormalTok{, }\FloatTok{2.6}\NormalTok{)) }\SpecialCharTok{+}
  \FunctionTok{theme}\NormalTok{(}\AttributeTok{legend.position =} \StringTok{"none"}\NormalTok{)}

\NormalTok{legend }\OtherTok{\textless{}{-}} \FunctionTok{ggplot}\NormalTok{(sum\_d, }
                 \FunctionTok{aes}\NormalTok{(}\AttributeTok{x =}\NormalTok{ var\_impmiy, }\AttributeTok{y =}\NormalTok{ cov\_impmiy, }\AttributeTok{color =}\NormalTok{ p\_miss\_1)) }\SpecialCharTok{+} 
  \FunctionTok{geom\_point}\NormalTok{(}\AttributeTok{alpha =} \DecValTok{0}\NormalTok{) }\SpecialCharTok{+} 
  \FunctionTok{labs}\NormalTok{(}\AttributeTok{color =}\NormalTok{ latex2exp}\SpecialCharTok{::}\FunctionTok{TeX}\NormalTok{(}\StringTok{"$}\SpecialCharTok{\textbackslash{}\textbackslash{}}\StringTok{Pr(R\_X=1|Z=1)$"}\NormalTok{)) }\SpecialCharTok{+} 
  \FunctionTok{theme}\NormalTok{(}
    \AttributeTok{legend.position =} \StringTok{"bottom"}\NormalTok{,}
    \AttributeTok{plot.background =} \FunctionTok{element\_blank}\NormalTok{(),}
    \AttributeTok{panel.background =} \FunctionTok{element\_blank}\NormalTok{(),}
    \AttributeTok{axis.title.x =} \FunctionTok{element\_blank}\NormalTok{(),}
    \AttributeTok{axis.title.y =} \FunctionTok{element\_blank}\NormalTok{(),}
    \AttributeTok{axis.text.x =} \FunctionTok{element\_blank}\NormalTok{(),}
    \AttributeTok{axis.text.y =} \FunctionTok{element\_blank}\NormalTok{(),}
    \AttributeTok{axis.ticks =} \FunctionTok{element\_blank}\NormalTok{(),}
    \AttributeTok{panel.grid =} \FunctionTok{element\_blank}\NormalTok{(),}
    \AttributeTok{plot.margin =} \FunctionTok{unit}\NormalTok{(}\FunctionTok{c}\NormalTok{(}\SpecialCharTok{{-}}\DecValTok{1}\NormalTok{, }\SpecialCharTok{{-}}\DecValTok{1}\NormalTok{, }\DecValTok{1}\NormalTok{, }\DecValTok{1}\NormalTok{), }\StringTok{"lines"}\NormalTok{)}
\NormalTok{  )}

\FunctionTok{ggsave}\NormalTok{(}\StringTok{"manuscript\_files/figure{-}pdf/legend.png"}\NormalTok{, }
       \AttributeTok{plot =}\NormalTok{ legend, }\AttributeTok{width =} \DecValTok{8}\NormalTok{, }\AttributeTok{height =} \FloatTok{0.5}\NormalTok{, }\AttributeTok{units =} \StringTok{"in"}\NormalTok{)}

\NormalTok{knitr}\SpecialCharTok{::}\FunctionTok{include\_graphics}\NormalTok{(}\StringTok{"manuscript\_files/figure{-}pdf/legend.png"}\NormalTok{)}
\end{Highlighting}
\end{Shaded}

\hypertarget{appendix-c---bayesian-analogue}{%
\section*{Appendix C - Bayesian
Analogue}\label{appendix-c---bayesian-analogue}}
\addcontentsline{toc}{section}{Appendix C - Bayesian Analogue}

To elaborate as to why the stochastic imputation using the outcome is
imitating a fully Bayesian process, we outline the Bayesian analogue of
the regression model with a covariate that has missing values. It is
common to consider a Bayesian model as having three stages. The first
stage is the model for the outcome variable, which is exactly the linear
regression model, \(Y = \beta_0 + \beta_1 X + \varepsilon_Y,\) where
\(\varepsilon_Y\) is normally distribution with mean zero and variance
\(\sigma^2\). The second stage is often referred to as the process
model, and in this setting specifies the imputation model,
\(X = \alpha_0 + \alpha_1 Z + \varepsilon_X,\) where \(\varepsilon_X\)
is normally distributed with mean zero and variance \(\tau^2\). Note
that these first two stages reflect our beliefs about the outcome model
and the missingness mechanism, and this model specification applies for
either frequentist or Bayesian inference.

In the Bayesian setting, we must add another layer to our model to
specify the prior distributions of our remaining parameters, say
\(\boldsymbol{\theta} = (\beta_0, \beta_1, \sigma^2, \alpha_0, \alpha_1, \tau^2)^T\)
. It is common in regression models to choose noninformative priors that
are flat on the real line for regression coefficients and ones that are
proportional to the inverse of the variance for the variance parameters.
Bayesian inference is performed using the posterior distribution, or the
distribution of all random, unknown quantities conditional on all
observed data. In this case, that is the distribution of
\(\boldsymbol\theta, \mathbf{X}_{mis}\) conditional on
\(\mathbf{Y}, \mathbf{X}_{obs}, \mathbf{Z}.\) We can simulate from the
posterior distribution using Gibbs sampling, which requires iteratively
sampling from the full conditional distributions. More specifically, we
first simulate from the conditional distribution of
\(\boldsymbol\theta\) given
\(\mathbf{Y}, \mathbf{X}_{obs}, \mathbf{Z}\), and then we use that value
of \(\boldsymbol\theta\) and simulate from the conditional distribution
of \(\mathbf{X}_{mis}\) given
\(\boldsymbol\theta,\mathbf{Y}, \mathbf{X}_{obs}, \mathbf{Z}\). Then we
repeat this process many times. It can be shown that the full
conditional distribution for \(X_{mis}\) is normally distributed with
mean of the form \(\gamma_0 + \gamma_1 Y + \gamma_2 Z\), where
\(\gamma_0, \gamma_1, \gamma_2\) and the variance are functions of
\(\boldsymbol\theta\). In fact, for a more general model, the mean will
be linear in the outcome and in all other covariates that are in the
outcome model, in addition to all covariates in the imputation model.
This motivates the form of the stochastic imputation using the outcome
setting. While that approach itself is not Bayesian, the requirement of
congeniality essentially means the model has a fully Bayesian analogue.

It is important to note that this model does not assume \(X\) depends on
\(Y\) in the imputation model. In fact, a Bayesian approach prohibits
modeling \(Y\) based on \(X\) and \(X\) based on \(Y\) as that is
cyclical and Bayesian models must correspond to directed acyclic graphs.
However, in the Bayesian setting inference is then performed conditional
on \(Y\), hence the involvement with the outcome when imputing missing
values of the covariate.

\end{document}